\documentclass[aps,groupaddress,nofootinbib]{revtex4}

\usepackage{slashed}
\usepackage{braket}
\usepackage{graphicx}
\usepackage{amsmath}
\usepackage{amssymb}
\usepackage{epsfig}
\usepackage{hhline}
\usepackage{soul}
\usepackage{tabularx,pifont,multirow}
\usepackage{enumitem}
\usepackage{colordvi}
\usepackage{xcolor}

\newcommand{\be}{\begin{equation}}
\newcommand{\ee}{\end{equation}}
\newcommand{\bea}{\begin{eqnarray}}
\newcommand{\eea}{\end{eqnarray}}

\newcommand{\nn}{\nonumber\\}

\newcommand{\nt}[1]{{\textcolor{black}{#1}}}
\newcommand{\gnote}[1]{{\textcolor{black}{#1}}}
\newcommand{\nicktext}[1]{{\textcolor{black}{#1}}}
\newcommand{\tmop}[1]{\ensuremath{\operatorname{#1}}}

\begin{document}
\preprint[\leftline{KCL-PH-TH/2020-{\bf 17}}

%

\title{ \bf
\boldmath
\Large     Dynamical Majorana Neutrino Masses and Axions  I} 


\author{Jean Alexandre, Nick E.~Mavromatos} 

\vspace{0.1cm}
\affiliation{Theoretical Particle Physics and Cosmology Group, Department of Physics, King's College London, Strand, London WC2R 2LS, UK}

\author{Alex Soto}
 
\affiliation{Instituto de Fisica, Pontificia Universidad Catolica de Chile
\\Vicu$\tilde n$a Mackenna 4860, Santiago 7820436, Chile, and \\ Theoretical Particle Physics and Cosmology Group, Department of Physics, King's College London, Strand, London WC2R 2LS, UK
}


\bigskip 

\begin{abstract}

\begin{center}
\textbf{Abstract }
\end{center}

We discuss dynamical mass generation for fermions and (string-inspired) pseudoscalar fields (axion-like particles (ALP)), in the context of effective theories containing Yukawa type interactions between the fermions and ALPs.  We discuss both Hermitian and non-Hermitian Yukawa interactions, which are motivated in the context of some scenarios for radiative (anomalous) Majorana sterile neutrino masses in some effective field theories. The latter contain shift-symmetry breaking Yukawa interactions between sterile neutrinos and ALPs.  Our models serve as prototypes for discussing dynamical mass generation in string-inspired theories, where ALP potentials are absent in any order of string perturbation theory, and could only be generated by stringy non perturbative effects, such as instantons. The latter are also thought of as being responsible for generating our Yukawa interactions, which are thus characterised by very small couplings.
We show that, for a Hermitian Yukawa interaction,  there is no (pseudo)scalar dynamical mass generation, but there is fermion dynamical mass generation, provided one adds a bare (pseudo)scalar mass. The situation is opposite for an anti-Hermitian Yukawa model: there is (pseudo)scalar dynamical mass generation, but no fermion dynamical mass generation.
In the presence of additional attractive four-fermion interactions, dynamical fermion mass generation can occur in these models, under appropriate conditions and range of their couplings.

\end{abstract}
\maketitle

\section{Introduction and Motivation \label{sec:intro}}

The mass of light active neutrinos, as evidenced by oscillations, points already towards physics beyond the Standard Model (SM), 
given that neutrino masses cannot be generated by SM-Higgs-like
Yukawa couplings, due to the absence of a right-handed neutrino in the SM spectrum. 
To date, the most widely accepted mechanism for generating light
neutrino masses is the see-saw~\cite{seesaw}, which necessitates
the Majorana nature of the light (active) neutrinos and the presence
of heavy right-handed Majorana partners of mass $M_R$, which is much
higher than the lepton or quark mass scale.\footnote{It should be mentioned, for completeness, that other more exotic approaches to the 
origin of neutrino masses, involving violation or modification of the Lorentz symmetry, do exist in the current literature, 
for instance in the so-called very special relativity framework~\cite{Cohen:2006ky}. In our approach here we maintain Lorentz invariance.}

A rather novel scenario was presented in ref.~\cite{mp} according to
which a \emph{radiative mechanism} for gauge-invariant mass generation
of chiral fermions has been proposed, which utilises global gravitational anomalies~\cite{anomalies}
in string-inspired effective field theories. The mechanism for generating light neutrino masses is triggered  by the 
existence of pseudoscalar fields (axion-like particles (ALP), from now called axions for brevity), that are abundant in (the moduli sector of) string models~\cite{arvanitaki}, and heavy right-handed fermions whose
masses are generated not by spontaneous symmetry breaking but, {\it radiatively}, as a consequence of (potentially non-perturbatively-induced)  shift-symmetry breaking Yukawa interactions with the axions. These interactions, combined with the gravitational anomalies, which the sterile neutrino couple to, generate a mass for the sterile neutrino, without traditional symmetry breaking, which is independent of the details of the axion potential. Such a mass is then communicated to the light (SM) neutrino sector via the traditional Higgs portal interaction terms connecting the sterile neutrino to Higgs and SM leptons. 

\gnote{The current article proposes an alternative way to generate dynamically a sterile-neutrino mass.}
In what follows, we shall  consider the Yukawa axion-sterile-fermion interaction 
through Schwinger-Dyson (SD) equations, and shall study the dynamical generation of masses for {\it both}, sterile fermions and axions $a(x)$, 
under certain conditions. Specifically,  as we shall  demonstrate in this work, dynamical mass generation for {\it both} fermion and (pseudo)scalar 
fields, with masses proportional to the product of the Yukawa coupling and the Ultra-Violet (UV) cutoff of the effective theory, 
is possible only in the presence of additional attractive four-fermion interactions. As we shall discuss in section \ref{sec:maj}, 
in the context of Majorana right-handed neutrinos of relevance to the work of \cite{mp}, when expressed in terms of the Majorana neutrino field, 
$\psi^M = \psi_R^C + \psi_R $, such four-fermion interactions read
 \begin{align}\label{4feff}
 -\frac{1}{2\,f_4^2} \Big(\overline {\psi^M}\, \gamma^5 \, \psi^M\Big)^2~,
\end{align}
while, as can be seen from Eq.\eqref{bacoupl2} below, the axion-right-handed neutrino terms can be written as 
\begin{align}\label{axmaj}
i \lambda \,  a(x) \, \overline{\psi^M }\, \gamma^5 \, \psi^M~. 
\end{align}
In the framework of strings, one may consider several axion fields $a_j(x)$, with different masses, which, as mentioned above, 
although do not play any role in the radiative mass of \cite{mp}, nevertheless  can couple to the Majorana fermions with terms of the form 
\eqref{axmaj}, but with different, in general, Yukawa couplings $\lambda_j$. 
Upon considering ultra-heavy axions, and integrating them out, in the context of the effective low-energy field theory,  one obtains, 
from the relevant super-massive-axion-exchange graphs, attractive interaction terms of the form \eqref{4feff}, with appropriate couplings 
\begin{align}\label{f4eff}
f_4^2 \propto \frac{M_j^2}{|\lambda_j^2|}~, 
\end{align}
which exist in the effective low-energy Lagrangian, in  to the interactions \eqref{axmaj} of (light or zero-bare-mass) axion fields $a(x)$.
 
 In our analysis in this work we shall allow the $a(x)$ field to have in general a non-zero bare mass.
As we shall demonstrate below, in the case of Hermitian Yukawa couplings, there are novel 
non-perturbative solutions  associated with the 
generation of light masses for both the axion and scalar fields, provided a small bare mass for the axion exists, 
proportional to $|\lambda  |\, \Lambda$, $|\lambda | \ll 1$, where $\Lambda$ is the ultraviolet cutoff of the theory. 
On the other hand, in the case of non-Hermitian Yukawa couplings only a pseudoscalar dynamical mass is generated, 
since energetics arguments prevent the generation of a fermion mass.

We shall also specify the conditions under which four-fermion terms of the form \eqref{4feff} will induce masses for the fermions 
and/or pseudoscalars, much smaller than the UV cutoff, or order $\sim \lambda \, \Lambda$, $|\lambda | \ll 1$. 
For this, it is necessary that the effective four-fermion coupling $f_4 $ is of order $\Lambda$, which is consistent with \eqref{f4eff}, 
for an appropriate range of the masses $M_j$. Phenomenologically, such self interactions,  among particles that could play the role of  
dark matter components (like sterile neutrinos and axions in our model), might be useful in discussing the growth of structure in 
the Universe and in particular the structure of galaxies, thereby offering ways to alleviate current tensions between the conventional 
$\Lambda$CDM-based simulations of galactic structure and observations~\cite{galaxy}. 

We also mention that four-fermion interactions of the form discussed here have also been considered in purely fermionic models in 
\cite{bender4f}, from \nt{a Parity-Time-reversal(PT)}-symmetry point of view, assuming, though, bare masses for the fermions. In our work, as already mentioned, 
we shall be concerned with the role of such interactions on assisting dynamical mass generation in theories involving physically motivated 
interactions between (pseudo)scalars and fermions, which has not been discussed in \cite{bender4f}, and in this sense our perspective is different. \nt{We shall also employ non-Hermitian Yukawa interactions, which will be physically motivated, in the context of ref.~\cite{mp}, in the next section}.

\nt{The structure of the article is as follows: in the next section, \ref{sec:straxion} we place our models in the more general context of 
stringy axion physics, and contrast the properties of our stringy axions from those of QCD axions. This section is intended for the general reader, 
so as to appreciate the difference of our models from other axion models in the contemporary literature.  
In section  \ref{sec:SD}}, we discuss the SD formalism for dynamical mass generation of scalar and fermion fields in prototype models, 
involving Dirac fermions. We employ both Hermitian and non-Hermitian Yukawa couplings, and show that for a Hermitian Yukawa interaction,  
there is no scalar dynamical mass generation, but there is fermion dynamical mass generation, provided a bare mass for the scalars is present. 
For an anti-Hermitian Yukawa model, on the other hand, there is (pseudo)scalar dynamical mass generation, but no fermion dynamical mass generation. 
In the presence of attractive four-fermion interactions, for an appropriate range of their couplings, dynamical fermion mass generation can occur 
in the non-Hermitian model, but also in the Hermitian model in  the presence of a bare (pseudo)scalar mass.
In section \ref{sec:maj} we discuss the extension of these results to the case of Majorana fermions, which has motivated the present study. 
The results are similar to the Dirac fermion case. 
Finally section \ref{sec:concl} contains our conclusion and plans for the future. 
Technical aspects of our SD approach, in both Hermitian and non-Hermitian models, are discussed in Appendix \ref{sec:appA}. 

\section{Review of the string-inspired model\label{sec:straxion}}

The effective field theory considered in \cite{mp} couples the Kalb-Ramond (KR) \emph{gravitational axion} field, 
$b(x)$ appearing in the fundamental massless gravitational multiplet of strings~\cite{string} after compactification to four space-time dimensions, 
with other axion fields (ALPs), $a_i(x)$, $i=1, \dots n$, that arise in string moduli~\cite{arvanitaki}. The coupling is provided  
by kinetic mixing Lagrangian terms, with coefficients $\gamma_i$, $i=1, \dots n$:
\begin{align}\label{mixing}
\mathcal L_{\rm mix}^{\rm kinetic}  = \gamma_i \, \partial_\mu b(x) \, \partial^\mu a_i(x), \quad i=1, \dots n.
\end{align}
 Diagonalising the axion kinetic terms, by appropriately redefining the fields, we arrive at an effective field theory for the axion fields $a_i(x)$, 
 by effectively decoupling the $b(x)$ field from the matter sector. For our purposes, it suffices to consider only one axion field $a(x)$. We refer the reader to \cite{mp} for the multiaxion field case. The pertinent, \nt {for our purposes, parts of the} effective action are (we work with metric signature conventions $(+, -, -, -)$ throughout)
\begin{eqnarray} \label{bacoupl2}
 \mathcal{S} & =  \int d^4 x \sqrt{-g} \, \Big[
\frac{1}{2} \Big(1- \gamma^2
\Big) \, (\partial_\mu a(x))^2 + i\, \overline \psi_R\slashed{\nabla} \, \psi_R
+ \lambda i a(x)\, \Big( \overline{\psi}_R^{\ C} \psi_R - \overline{\psi}_R
\psi_R^{\ C}\Big) \Big] + \dots\;, 
\end{eqnarray}
\nt{where $g$ is the determinant of the metric tensor,  $\psi_R$ is a right-handed (chiral) fermion, 
which plays the role of the sterile neutrino, $\psi_R^{\ C} = (\psi_R)^C$ is its (Dirac) charge-conjugate, and $\nabla_\mu $ denotes the (torsion-free) gravitational covariant derivative. The $\dots$ denote terms which are of no direct relevance for our purposes in this article, except the 
gravitational anomaly terms, coupling the axion to the Riemann-space-time curvature tensor, which will be relevant in a way to be specified below.}

The Yukawa coupling $\lambda$ may be due to non-perturbative string instanton effects, and as such is expected to be generically small. These
terms \emph{break explicitly} the shift symmetry $a(x) \to a(x) + c$, in the same spirit as the instanton-generated potential for the axion fields.  From \eqref{bacoupl2} we
observe that, for real $\gamma$,  
the approach is only valid for
\begin{align}\label{gammaregion}
\gamma < 1\ ,  \qquad \gamma \in {\tt R}, 
\end{align}
otherwise the axion field would appear as a ghost, \emph{i.e.} with
the wrong sign in its kinetic term, which would indicate an
instability of the model and trouble with unitarity. This is the only restriction of the
parameter $\gamma$ in the work of \cite{mp}.

\begin{figure}[t]
 \centering
  \includegraphics[clip,width=0.40\textwidth,height=0.15\textheight]{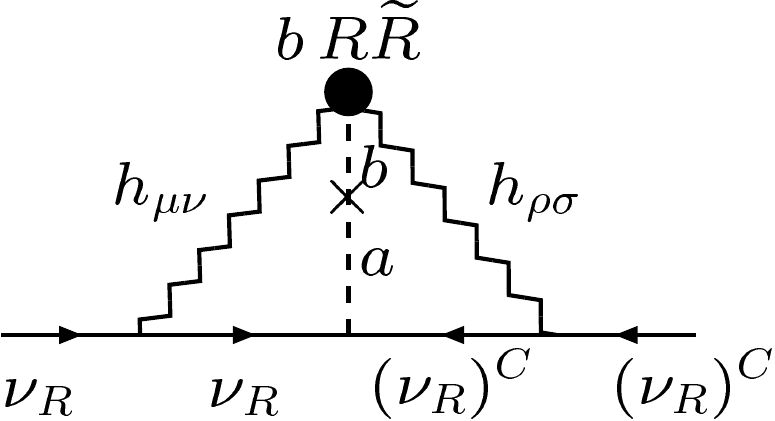} 
\caption{\it Typical Feynman graph giving rise to anomalous 
 Majorana mass generation for the right-handed fermions, corresponding to sterile neutrinos $\psi_R \equiv \nu_R$ in ref.~\cite{mp}.  The black circle denotes the operator $b(x)\,
  R_{\mu\nu\lambda\rho}\widetilde{R}^{\mu\nu\lambda\rho}$, where $b(x)$ denotes the KR (gravitational) axion and $a(x)$ an ALP (also called axion from now on, for brevity). The wavy lines are gravitons $h_{\mu\nu}$. Dashed lines are axions. The b(x)-a(x) kinetic mixing is indicated on the axion line by a cross ``$\times$''.}\label{fig:anom}
\end{figure}
However, if one is prepared to use non-Hermitian Lagrangians, which will be partly the topic of our current work, 
one may relax the reality condition for $\gamma$ and also allow for purely imaginary values
\begin{align}\label{ig}
\gamma = i \tilde \gamma, \qquad \tilde \gamma \in {\tt R}
\end{align}
We shall come back to this important point later on.

For the moment, we note that, in either case, we may redefine the axion field so as to appear with a
canonically normalised kinetic term:
\begin{eqnarray} \label{bacoupl3} 
\mathcal{S}_a \!&=&\! \int d^4 x
    \sqrt{-g} \, \Big[\frac{1}{2} (\partial_\mu a(x) )^2  +  i\,  \, \overline \psi_R \slashed{\nabla} \, \psi_R 
+ \frac{i\, \lambda}{\sqrt{1 - \gamma^2}} \, 
a(x)\, \Big( \overline{\psi}_R^{\ C} \psi_R - \overline{\psi}_R
\psi_R^{\ C}\Big) \Big] + \dots\;.
\end{eqnarray}

The reader should recall that, in a Minkowski or  Friedman-Lema$\hat{\rm i}$tre-Robertson-Walker (FLRW) space-time backgrounds, 
the gravitational anomaly term {\it vanishes}, but this is not the case for generic 
quantum metric fluctuations about such a background, or gravitational-wave type classical fluctuations~\cite{bams,aps}. 
\nt{Such graviton fluctuations couple to the sterile neutrinos in the model of \cite{mp}, and are responsible for an anomalous 
(radiative) Majorana mass generation for such fields, through the mechanism shown in Fig.~\ref{fig:anom}.} 
Adopting the framework of \cite{Donoghue:1994dn}, one may  estimate  the  two-loop  Majorana neutrino mass~\cite{mp}: 
\begin{align}\label{massR}
M_R \sim \frac{\sqrt{3}\, \lambda\, \gamma\, c_1\, \kappa^5 \Lambda^6}{49152\sqrt{8}\,
\pi^4 (1 - \gamma^2 )}\; ,  
\end{align}
\nt{where $c_1$ denotes the number of chiral fermion fields in the model, and} $\Lambda$ is an Ultra-Violet (UV) momentum cutoff. In an UV 
complete theory,  such as  strings, $\Lambda$  and $M_P$ (or, equivalently, $\kappa^{-1}$)  are related~\cite{mp}. 
For a generic quantum gravity model, independent of string theory, one may use simply  $\Lambda \sim \kappa^{-1}$. 
We stress that the so-induced sterile fermion mass \eqref{massR} is {\it independent} of the axion-$a(x)$ potential~\cite{mp}. 

An important comment we would like to make concerns the evasion of the constraint \eqref{gammaregion}, 
via the complexification procedure \eqref{ig}.  In such a case one is effectively working with {\it non-Hermitian} Hamiltonians but connected to the so-called PT-symmetric framework~\cite{bender} \nt{(with P denoting parity (spatial-reflection) operation, and T the time reversal operation, as already mentioned in the introduction}), stemming from the effective action \eqref{bacoupl3}, upon the {\it simultaneous} complexification of the parameters $\gamma, \lambda$, which now become purely imaginary 
\begin{align}\label{complex}
\gamma \rightarrow i \tilde \gamma, \quad \lambda \rightarrow i \tilde \lambda, \qquad \tilde \gamma, \, \tilde \lambda \in {\tt R}.
\end{align}
Indeed, such a procedure results in non-Hermitian axion-sterile neutrino Yukawa couplings. Had one had a scalar field in such interactions, one would obtain PT-symmetric Hamiltonians~\cite{bender} of the type discussed in \cite{ptneutrino}, which have been argued to be consistent field theories describing phenomenologically relevant neutrino oscillations. In our axion case, the non-Hermitian Yukawa interaction is PT-odd. Nonetheless, as discussed in \cite{AM}, such interactions can lead to real energies in a certain regime of their parameters, thus making a connection with PT-symmetric systems~\cite{BJR}.  
\nicktext{In fact, according to the discussion in \cite{mann}, the existence of real energies in systems with
non-Hermitian Hamiltonians is guaranteed if one has an underlying anti-linear symmetry, which could be more general than PT.  In our case such an antilinear symmetry is the CPT symmetry~\cite{AM}.} 
In \cite{AM} we also demonstrated the consistency of such non-Hermitian models with Lorentz invariance (including  also improper Lorentz transformations), as well as unitarity~\cite{AMS}.

In our case, the axion $a(x)$ coupling to the gravitational anomaly constitutes another non-Hermitian contribution, which, however, as explained above, if one ignores graviton or gravitational-wave fluctuations, vanishes for FLRW space-times. 
Graviton fluctuations, though, are important for radiative generation of sterile neutrino mass~\cite{mp}, as we have discussed above. Thus,  
formally, such a non-Hermitian approach extends the range of the radiatively $\tilde \gamma$ to all the real axis, 
leading to a {\it real} sterile-neutrino mass \eqref{massR} :
\begin{align}\label{massR2}
M_R \sim - \frac{\sqrt{3}\, \tilde \lambda\, \tilde \gamma\, c_1\,  \kappa^5 \Lambda^6}{49152\sqrt{8}\,
\pi^4 (1 + \tilde \gamma^2 )}\; .
\end{align} 
Notice that the sign of the fermion mass depends on the sign  of the product $\tilde \lambda\, \tilde \gamma$, and can always be chosen to be positive, although for fermions, unlike bosons,  such a sign is not physically relevant. 

At this point we would like to comment that in the non-Hermitian case, for consistency, the classical axion field equation of motion, stemming from \eqref{bacoupl3}, yields (upon considering FLRW space-time backgrounds, ignoring the effects of graviton fluctuations):
\begin{align}\label{axion}
\Box a(x) =  \frac{\tilde \lambda}{\sqrt{1 - \gamma^2}} \, \Big(\overline{\psi}_R^{C} \psi_R - \overline{\psi}_R \, \psi_R^{C} \Big) + \dots ~,
\end{align}
with $\Box a(x)=g^{\mu\nu} \nabla_\mu \partial_\nu a(x)$ the torsion-free gravitationally-covariant D' Alembertian, and the $\dots$ denote the gravitational anomaly terms. If we ignore the latter, then we observe, that, as a result of 
the non-Hermitian nature of the right-hand side of the above equation, the necessary solution is that of a free axion field, and vanishing vacuum expectation values (classical configurations) for the spinors~\cite{AM}:
\begin{align}\label{axion2}
\Box a (x) =0, \qquad  \langle 0 | \overline{\psi}_R^{C} \psi_R |0 \rangle = \langle 0| \overline{\psi}_R \, \psi_R^{C} |0\rangle =0.
\end{align}
A trivial solution for the fermions, in such non-Hermitian models, would be that for vanishing background configurations, but with non-trivial quantum fluctuations, leading to dynamical mass generation for both the sterile neutrinos and the axion field.  In this work we shall also discuss more complicated non-Hermitian models, leading to non-trivial classical solutions for the fermions and scalar fields.

\nicktext{Let us make some important comments here, concerning the difference of our axion / pseudoscalar fields we shall consider in the prototype model discussed here, 
and in its sequel paper~\cite{MS2}, from standard QCD axions, ${\mathcal A}(x)_{\rm QCD}$. In the latter case, there is a non-perturbative (instanton) generated potential, due to the anomalous coupling of the axion to gluons, which yields a periodic potential for the QCD axions of the (generic) form 
\be\label{qcdaxion}
V_{\rm QCD} ({\mathcal A}(x)_{\rm QCD})  = \Lambda^4_{\rm QCD}\, \Big[1 - {\rm cos}\Big(\frac{{\mathcal A}_{\rm QCD}(x))}{f_a}\Big)\Big],
\ee
where $\Lambda^4_{\rm QCD}$ is the QCD scale, 
and $f_a$ is the QCD axion coupling constant, with dimensions of mass, which determined the 
scale of the anomalous breaking~\cite{QCDaxions} of the Peccei-Quinn symmetry~\cite{PQ}. The resulting mass of the QCD
axion is determined by  
\be\label{axionmass}
m_{\mathcal A_{\rm QCD}} = 6 \times 10^{-10} \, \Big(\frac{10^{16} \, {\rm GeV}}{f_a}\Big)\, {\rm ev}.
\ee
In addition to its couplings to gluons, the QCD axion can also couple anomalously (via chiral anomalous loop graphs) to photons, and could also have {\it derivative couplings} to fermions. 
Experimental searches exclude $f_a < {\mathcal O}(10^9)~{\rm GeV}$.}

\nicktext{As already mentioned, the pseudoscalar/ALP $a(x)$, $b(x)$ ({\it cf.} \eqref{mixing}), we consider in this work and its sequel \cite{MS2}, 
as well as in \cite{mp}, which inspired the current articles, are string theory axions that populate the string {\it Axiverse}~\cite{arvanitaki}. Perturbatively, such axions do not have any potential in perturbative string theories, as
they stem from appropriate flux fields in string theory, which obey certain gauge symmetries. For instance, the Kalb-Ramond (KR) gravitational axion $b(x)$ stems from the field strength  $H_{\mu\nu\rho} = \partial_{[\mu }B_{\nu\rho]}$ of the spin-1 antisymmetric tensor field $B_{\mu\nu}=-B_{\nu\mu}$ (where $[\dots]$ denote antisymmetrisation of the indices), 
which in four-dimensional space-times is dual to the derivative of $b(x)$: 
\be\label{H}
H_{\mu\nu\rho} \propto \varepsilon_{\mu\nu\rho\sigma} \, \partial^\sigma b(x),
\ee
with $ \varepsilon_{\mu\nu\rho\sigma}$ the gravitationally covariant Levi-Civita antisymmetric tensor.
 The field strength $H$ 3-form is invariant under a gauge U(1) symmetry 
 \be
 B_{\mu\nu} (x)  \to B_{\mu\nu} (x) + \partial_\mu \theta_\nu (x) - \partial_\nu \theta_\mu (x), 
 \ee
 and thus the (perturbative) closed-string sector effective action, which is invariant under
 this U(1) symmetry~\cite{string}, depends only on $H_{\mu\nu\rho}$ and its derivatives, leading to the absence of a KR-axion-$b(x)$ potential. Similar gauge symmetries characterise higher-order forms which, upon compactification to four space-time dimensions give rise to ALPs without potentials~\cite{arvanitaki} in  any order in string perturbation theory.}

Such gauge symmetries may or may not break by non-perturbative effects in string/brane theory. 
Their breaking depends on specific stringy/brane backgrounds, and it is not a generic effect~\cite{arvanitaki}. Even if a non-perturbative potential is generated for such string theory axions, $a(x)$, its mass $m$ and the corresponding axion constant $f_a$ are two parameters, which are not related by the QCD axion relation \eqref{axionmass}. 
There may be several non-perturbative effects, generating stringy ALP potentials at different scales $\Lambda_i$, such that the overall potential is given by 
\be\label{axionstring}
\sum_{i} \Lambda_i^4 \, \mathcal U(a),
\ee
where $\mathcal U(a)$ is a periodic function with a period $2\pi$. The scales 
$\Lambda_i$ can be taken to be 
\be\label{liscales}
\Lambda_i ^4= \Lambda^4 \, e^{-S_i} ,
\ee
where $\Lambda$ is some UV cutoff, which in string theory may be taken to be the string scale, and $S_i$ appropriate scales representing instanton actions. The corresponding axion masses are determined by 
\be\label{axionmassstring}
m_i = \frac{\Lambda_i^2}{f^{\rm str}_a}\,.
\ee
where $f^{\rm str}_a$ are appropriate stringy axion coupling constants with dimensions of mass. 
In the model of \cite{mp} we also considered non-perturbative shift-symmetry breaking Yukawa couplings of stringy ALPs to 
right-handed sterile neutrino fields, as an additional non-perturbative effect. 
In what follows, we may consider such effects as being generated at energy scales higher than 
any of the energy scales $\Lambda_i$ which generate axion potentials. The question we are trying to answer is here, and in the sequel paper \cite{MS2}, is whether such interactions can generate 
fermion but also ALP masses. Our Yukawa couplings will be taken to be extremely weak, being due to non-perturbative stringy effects. 
\nt{Our UV cutoff $\Lambda$ will be taken to be of order of the (four-dimensional) Planck scale, $\Lambda \sim M_P \sim 1.22 \times 10^{19}$~GeV, 
and in the model of \cite{mp}, the KR-$b(x)$-axion constant is also of that order~\cite{mp}:\begin{align}\label{fb}
f_b \equiv (3\kappa^2/8)^{-1/2} = \frac{M_P}{\sqrt{3\pi}}. 
\end{align}}

\nicktext{In this sense, the models we are going to consider are relevant to the very early Universe, for instance in models like the string-inspired models of a ``running cosmological vacuum'' discussed in \cite{bams}, where the effective string-inspired field theory describing the early Universe is assumed to consist only of fields appearing in the massless gravitational  multiplet of strings, including stringy ALPs stemming from string moduli fields. 
Our approach on generating masses for right-handed fermions in such models to be discussed below and in \cite{MS2}, is therefore relevant for such approaches, given the connection 
of massive right handed primordial neutrinos in the background of stringy ALPs for leptogenesis, according to the scenarios presented in \cite{decesare}. The model we envisage, at least in the context of the running vacuum approach of \cite{bams}, is that stringy ALPs, like KR axions, without any potential, couple to primordial right-handed neutrinos during the early universe, and the question we ask is whether dynamical mass can be generated as a result of such interactions.\footnote{\nicktext{Nonetheless, for the models with Hermitian Yukawa interactions, we shall need bare axion masses in order to generate non-perturbatively right-handed fermion masses. As we shall discuss later on ({\it cf.} \eqref{cosine}), such masses can be generated by the same 
non-perturbative physics, deep in the Quantum Gravity regime, that generates the shift-symmetry breaking Yukawa interactions in \eqref{bacoupl2}, with a stringy-instanton-induced $\lambda \ll 1$. For such weak Yukawa couplings, self interactions of the axion field can be ignored in the early Universe phase, leaving us with only a bare mass term for the axion, proportional to $\lambda$, as the dominant term in the quantum-gravity-induced potential. The dynamical mass for the axion (and fermion) are much smaller than the bare mass (\eqref{SD1cb}).}}
In the model of \cite{bams}, chiral matter, including right-handed neutrinos, as well as gauge/radiation fields are generated at the late stages/exit of inflation, and when mass is generated for such neutrinos, e.g. through the Yukawa interaction mechanisms we discuss here, they can lead to leptogenesis  during the radiation era, as a result of their decay in CPT violating backgrounds of the KR axions~\cite{decesare}, which survive undiluted the exit from inflation. The stringy axions in the model acquire their potential through anomalous coupling with gauge fields during the radiation phase, which follows the inflationary period.}

\nicktext{In general, the inclusion of scalar potentials in our study, of relevance to subsequent eras after inflation for the string models of \cite{bams},  is a complicated task, and depends on the 
details of the underlying string model. 
Above, we have explained the reasons why this is not considered in the present work and in \cite{MS2}. 
We also remark that our analysis below and in \cite{MS2} will be at zero temperature.
Such conditions might be relevant for a low temperature phase before reheating of the universe,
otherwise finite temperature studies become necessary. This falls beyond the scope of our prototype studies. 
As a further remark, we also note that in the context of string theory and its D-brane extensions~\cite{string}  there are detailed models by which stringy instantons yield masses for right-handed neutrinos~\cite{cvetic}. Embedding our models in such detailed constructions, and finding links between our approach and those models, remains an interesting avenue for research.} 

\nicktext{In addition to Hermitian hamiltonians, in our work here and in \cite{MS2}, we shall also study for the first time in the relevant literature, 
dynamical mass generation for non-Hermitian Yukawa axion-fermion couplings, which are consistent quantum theories within the PT-symmetry 
framework~\cite{AM,bender4f}. 
For such non-Hermitian theories, it is not yet known whether the generation of non-perturbative potentials for axions is possible.}

\section{Yukawa Interactions and Schwinger-Dyson Dynamical Mass generation  \label{sec:SD}}

We commence our analysis by considering the following prototype Lagrangian, involving a  (pseudo)scalar and a non-chiral (Dirac) fermion field. There will be no difference in our conclusions on dynamical mass generation if the field is scalar or pseudoscalar. Of course, from the point of view of the motivation for this work, as outlined in section \ref{sec:intro}, we are primarily interested in the pseudoscalar case. We commence our study with the Hermitian Yukawa interaction. Below we shall describe the basic results. Technical details are presented in the 
Appendix \ref{sec:appA}, where we discuss the SD equations for both Hermitian and non-Hermitian Yukawa interactions. 

\subsection{Hermitian Yukawa Interactions \label{sec:herm}} 
 
We consider a Dirac fermion and a real (pseudo) scalar field, with the conformal Lagrangian 
\be\label{model1}
L=\frac{1}{2}\partial_\mu\phi\partial^\mu\phi+\overline\psi i\slashed\partial\psi+i\lambda\phi\overline\psi\gamma^5\psi~.
\ee
For real $\lambda$, the Yukawa interaction is real, since $\overline\psi\gamma^5\psi$ is anti-Hermitian. \nicktext{For reasons explained above, we do not consider self interactions of the 
field $\phi$.} 

We ignore for our purposes in this section the gravitational background, and we work exclusively with a Minkowski space-time metric with signature $(+,-,-,-)$. 
If there is dynamical mass generation, one should take into account all the possible mass terms, which are
\be\label{dynmasses}
\frac{1}{2}M^2\phi^2~,~~m\overline\psi\psi~,~~i\mu\overline\psi\gamma^5\psi~,
\ee
where $M,\,m,\,\mu$ are real. 

As described in Appendix \ref{sec:hyi}, the SD equations for the scalar and fermion propagators, obtained by using standard field theoretic techniques~\cite{Peskin:1995ev}, read:
\bea\label{firstSD}
G_f^{-1}(k)-S_f^{-1}(k)&=&\lambda\gamma^5\int_p G_f(p)\Gamma^{(3)} (p,k) G_s(p-k)\\
G_s^{-1}(k)-S_s^{-1}(k)&=& - \mbox{Tr}\left\{\lambda\gamma^5\int_p G_f(p) \Gamma^{(3)} (p,k) G_f(p-k)\right\}~,\nonumber
\eea
where the index $s$ refers to the scalar and the index $f$ refers to the fermion. $G_{s,f}$ denote the dressed propagators, $S_{s,f}$ denote the bare propagators, 
$\Gamma^{(3)} (p,k)$ is the dressed vertex, and we abbreviated the four-momentum integrals by $\int_p \equiv \int  \frac{d^4 p}{(2 \pi)^4}$. 

We have
\be
S_f^{-1}(k)=-i \slashed k~~~~\mbox{and}~~~~S_s^{-1}(k)=-i k^2~,
\ee
and the lowest order approximation for the dressed propagators consists in allowing the dynamical generation of masses only, in which case we have
\be
G_f^{-1}(k)=- i (\slashed k-m-i\mu\gamma^5)~~~~\mbox{and}~~~~G_s^{-1}=-i (k^2-M^2)~.
\ee
To the lowest order approximation, one also neglects corrections to the vertex (\emph{rainbow} approximation), such that 
\begin{align}\label{vertex}
\Gamma^{(3)} (p,k) \simeq -\lambda\gamma^5~.
\end{align}

Some important remarks are due at this point. First, we stress that in our approximations, the fermion wave function renormalization is assumed to be 1. This is because, throughout this work, we are interested on very small Yukawa couplings, $|\lambda| \ll 1$, motivated by the considerations of ref.~\cite{mp}, where such Yukawa couplings are assumed to be induced non-perturbatively, through instanton effects. In this sense the ladder approximation 
considered in our approach suffices for discussing dynamical mass generation, for both scalar and fermions simultaneously, rather than discuss finer approximations. Such more elaborate approximations characterise the model of ref.~\cite{bashir}, which however is different from ours in the sense that
these authors consider a massless scalar with a quartic self interaction. In our case, we do not have such an interaction, but, as we shall discuss below, consistency requires that the (pseudo)scalar particle has a bare mass, proportional to the UV cut-off. \nicktext{In the context of the string-inspired model of \cite{mp}, which motivated our studies, the cutoff is taken to be the Planck scale $M_P$.}

An additional remark we would like to make concerns the 
consistency of \nt{Ward-Takahashi identities}, and therefore the form of the dressed vertex. 
This is relevant only when one considers the momentum-dependence of quantum corrections to the propagators, 
which is not the case of our study, as explained above. \nt{Ward-Takahashi} identities are satisfied in our approximation, 
up to higher orders in the small Yukawa coupling. In \cite{bashir} such quantum corrections have been discussed in connection  
with fermion mass generation only, in that model. The latter, however, is characterised by the presence of a critical coupling 
(common among the Yukawa and self interactions of the model), above which dynamical fermion mass takes place. Par contrast, 
in our model, as already mentioned, both scalar and fermion masses are dynamically generated for very small Yukawa couplings. 
\nicktext{The smallness the Yukawa couplings also imply that any fermion-loop generated (pseudo)scalar self-interactions, 
e.g. a quartic term, will be suppressed by higher powers of the Yukawa coupling, and as such will be subleading to the effects considered here.}

For vanishing external momenta, then, the SD equations read
\bea\label{SDbis}
i(m+i\mu\gamma^5)&=&-\lambda^2\gamma^5\int_p G_f(p)\gamma^5G_s(p)\\
i M^2&=&\mbox{Tr}\left\{\lambda^2\gamma^5\int_p G_f(p)\gamma^5 G_f(p)\right\}~.\nonumber
\eea
The momentum integrals are regulated with an UV  cut off, $\Lambda$, which also plays the role of mass scale in the system.
Dynamical mass occurs if the set of equations (\ref{SDbis}) has a non-trivial self-consistent solution for the masses (\ref{dynmasses}).

The details of constructing the system of appropriate SD equations are given in Appendix \ref{sec:hyi},
\gnote{where we use a cut-off regularisation. Unlike dimensional regularisation, this will allow us to take into account power-law divergences arising 
from four-fermion interactions, which will be introduced in the next subsection \ref{sec:4f}. 
The Pauli Villars regularisation would also show power-law divergences, but the cut-off regularisation employed in the present work is consistent with the original 
string-inspired model, where the Planck mass plays the r\^ole of a natural cut-off scale for the low-energy effective field theory.} 
The SD equations read:
\begin{align}\label{hSD1a}
  m + i\mu \gamma^5 = \frac{\lambda^2}{16 \pi^2} \frac{(m - i\mu \gamma^5)}{M^2 -
  m^2 - \mu^2} \left[ M^2 \ln \left( 1 + \frac{\Lambda^2}{M^2} \right) - (m^2
  + \mu^2) \ln \left( 1 + \frac{\Lambda^2}{m^2 + \mu^2} \right) \right]
\end{align}
\begin{align}\label{hSD2a}
  M^2 = -\frac{\lambda^2}{4 \pi^2} \left[ \frac{(\Lambda^2 + m^2 + 3 \mu^2)
  \Lambda^2}{\Lambda^2 + m^2 + \mu^2} - (m^2 + 3 \mu^2) \ln \left( 1 +
  \frac{\Lambda^2}{m^2 + \mu^2} \right) \right]
\end{align}

Splitting \eqref{hSD1a} in the part containing $\gamma^5$ and the part without it, we obtain the set of the SD equation for the fermion and (pseudo)scalar masses:
\begin{align}\label{24e}
  m = \frac{\lambda^2}{16 \pi^2} \frac{m}{M^2 - m^2 - \mu^2} \left[ M^2 \ln
  \left( 1 + \frac{\Lambda^2}{M^2} \right) - (m^2 + \mu^2) \ln \left( 1 +
  \frac{\Lambda^2}{m^2 + \mu^2} \right) \right]
\end{align}
\begin{align}\label{25e}
  \mu = - \frac{\lambda^2}{16 \pi^2} \frac{\mu}{M^2 - m^2 - \mu^2} \left[ M^2
  \ln \left( 1 + \frac{\Lambda^2}{M^2} \right) - (m^2 + \mu^2) \ln \left( 1
  + \frac{\Lambda^2}{m^2 + \mu^2} \right) \right]
\end{align}
\begin{align}\label{26e}
  M^2 = - \frac{\lambda^2}{4 \pi^2} \left[ \frac{(\Lambda^2 + m^2 + 3 \mu^2)
  \Lambda^2}{\Lambda^2 + m^2 + \mu^2} - (m^2 + 3 \mu^2) \ln \left( 1 +
  \frac{\Lambda^2}{m^2 + \mu^2} \right) \right]
\end{align}
The reader should recall that, for consistency of our SD approximations, we work in the small Yukawa coupling limit 
\begin{align}\label{yuklim}
|\lambda | \ll 1~.
\end{align}
This limit characterises all cases discussed in the current work.

We next proceed to solve the above equations. There are various cases, of physical relevance to our purposes, which we can consider.

\vspace{0.2cm}

\begin{itemize} 

\item{(i)} We first remark that, as can be seen from \eqref{24e}, \eqref{25e}, there is {\it no solution} with {\it both} $\mu \ne 0$ and $m \ne 0$. 

\item{(ii)} Second, we observe that, although the 
\emph{fermion} equations allow for trivial \emph{massless} solutions for the fermions, $\mu=m=0$, the (pseudo)scalar cannot dynamically a non-zero mass, since in that case:
\begin{align}\label{massaxion}
M^2 = - \frac{\lambda^2}{4\pi^2} \Lambda ^2  \, < \, 0~.
\end{align} 
In this case, since dynamical mass generation cannot occur, the only alternative mechanism for generating masses for the fermions would be the radiative mechanism, through gravitational anomalies (see Fig.~\ref{fig:anom}), described in section \ref{sec:intro}; this would lead to a fermion mass $m$ of the form \eqref{massR} (\eqref{massR2}) for the (non-)Hermitian Yukawa interaction case, which, as we have explained, is independent of the details of the axion potential, and hence its mass~\cite{mp}. 

\item{(iii)} We now seek solutions with $\mu=0$, $m \ne 0$. Notice that the vanishing of the chiral mass $\mu=0$ is consistent with the vanishing chiral condensate of the classical fermions in the non-Hermitian case, required by mathematical self consistency of the non-Hermitian model (see Appendix \ref{sec:appA}, \eqref{axioneom}, \eqref{cond}, and also \eqref{axion}, \eqref{axion2}). This restriction, of course, does not apply to the Hermitian case, but here, for simplicity, and in the spirit of our physical motivation, outlined in section \ref{sec:intro}, we do not consider chiral mass ($\mu$) generation for fermions when $m=0$ (the reader should recall from (i) that there is no solution for $\mu m \ne 0$).

On putting $\mu = 0$, we obtain the following SD equations:
\bea\label{SD1}
1&=&\frac{\lambda^2}{16\pi^2}\frac{1}{M^2-m^2}\left[M^2\ln\left(1+\frac{\Lambda^2}{M^2}\right)-m^2\ln\left(1+\frac{\Lambda^2}{m^2}\right)\right]\\
M^2&=- &\frac{\lambda^2}{4\pi^2}\left[\Lambda^2- m^2 \, \ln\left(1+\frac{\Lambda^2}{m^2}\right)\right]\nonumber~,
\eea
First, one can see that there is no scalar mass generation in the system, as follows from the 
second of the equations \eqref{SD1}, given  that $\Lambda^2 \, > \, m^2$ (the cut-off is assumed the highest 
mass scale in the problem). Upon setting $M=0$, we then obtain from \eqref{SD1};
\begin{align}\label{SD1b}
\Big(\frac{\Lambda}{m}\Big)^2 \simeq \ln \Big(1 +  \Big(\frac{\Lambda}{m}\Big)^2 \Big) \simeq \frac{16\, \pi^2}{\lambda^2}, \quad |\lambda| \ll 1.
\end{align}
These equalities are incompatible, so there is no dynamical fermion mass generation  if $M=0$. 

We now assume the existence of a {\it bare mass} for the scalar field,  $M_0 \ne 0$. In that case, it is the scalar SD 
equation from the system \eqref{firstSD} that is affected, since now $S_s^{-1}(k) = -i (k^2 - M_0^2)$.  The fermion SD equation remains as in the $M_0=0$ case.

We then arrive at the following SD system of equations (for $\mu=0$ that we adopt here, as mentioned earlier):
\bea\label{SD1c}
1&=&\frac{\lambda^2}{16\pi^2}\frac{1}{M^2-m^2}\left[M^2\ln\left(1+\frac{\Lambda^2}{M^2}\right)-m^2\ln\left(1+\frac{\Lambda^2}{m^2}\right)\right]\\
M^2&=& M_0^2 - \frac{\lambda^2}{4\pi^2}\left[\Lambda^2- m^2 \, \ln\left(1+\frac{\Lambda^2}{m^2}\right)\right]\nonumber~,
\eea
First, it is clear that we must have $M^2  < M_0^2$. Let us look for solutions $m \simeq M \ll \Lambda$. We have from 
\eqref{SD1c} 
\bea\label{SD1cb}
m^2 &\simeq&  \exp \Big(-\frac{16\, \pi^2}{\lambda^2}\Big) \, \Lambda^2, \quad |\lambda| \ll 1, \nonumber \\
M^2 &\simeq &  m^2 = M_0^2 - \frac{\lambda^2}{4\pi^2}\, \Lambda^2, \quad M_0^2 
= \frac{\lambda^2}{4\pi^2}\, \Lambda^2 + \exp \Big(-\frac{16\, \pi^2}{\lambda^2}\Big) \, \Lambda^2.
\eea
which indicates a non-perturbative (in  the Yukawa coupling $\lambda$) small dynamical fermion and (pseudo)scalar masses. 

\nicktext{The reader should notice that despite the presence of a bare (pseudo)scalar mass, proportional to the cutoff $\Lambda$, the resulting dynamical masses \eqref{SD1cb} for both (pseudo)scalar and 
fermion fields are indeed much smaller than the cutoff. Moreover, one may consider $\lambda \ll 1$ 
such that as $\Lambda \to \infty$, $\lambda \to \infty$, in such a way that the bare mass $M_0$ is finite. At any rate, as we have already mentioned, in the context of string-inspired models \cite{mp}, we consider here, $\Lambda$ is of the order of Planck mass scale $M_P$. According to our discussion above, this will lead to small but finite dynamical masses.}

An important comment we would like to make concerns the physical significance of the 
bare axion Mass $M_0$. One might have thought that the same non-perturbative physics that
generated the Yukawa coupling $\lambda$, is responsible for the generation of $M_0$. For example, one may think of some {\it transplanckian stringy} effects (which exist in string theory), 
deep in the stringy quantum gravity (QG) regime, which 
create,  in the early Universe, in addition to the Yukawa interaction, a non-perturbative 
axion-$\phi$ potential of the form
\be\label{transpl}
U_{\rm string} = \Lambda^4_{\rm QG} \, \Big(1 - {\rm cos}(\frac{\phi}{f_\phi})\Big)
\ee 
where 
\be\label{trans}
f_\phi =  c_1\, M_P/\lambda ,
\ee 
a transplanckian ``axion coupling constant'',
\footnote{\nt{Such axion constants can find a natural explanation in the string-inspired model of \cite{mp}, which may be taken as a prototype for the description of the dynamics of a primordial string-inspired Universe. To understand this, one may consider, as a toy mechanism, the coupling of the KR $b(x)$ field to primordial (non-Abelian) stringy gauge fields through the mixed (gravitational and gauge) anomalous interaction Lagrangian terms (in generic differential form notation, and refraining from writing explicitly the values of the anomaly coefficients $\mathcal C_{R,F}$, for brevity)  
\be\label{bff} 
\frac{1}{f_b} \, b(x)  \, \Big[ \mathcal C_R \, \mathbf R\, \wedge \widetilde{\mathbf R} \, - \mathcal C_F \, {\rm Tr}\Big({\mathbf F} \, \wedge \,\widetilde {\mathbf F}\Big)\Big],
\ee
with $f_b\sim M_P/\sqrt{3\pi}$ ({\it cf.} \eqref{fb}), $\mathbf F$ the non-Abelian gauge field strengths,  $\mathbf R$ the Riemann-space-time-curvature two-form, 
the tilde over a form $\mathbf A$, $\widetilde{\mathbf A}$, denoting its dual, the $\wedge$ referring to exterior products of differential form calculus,  and ``Tr'' representing gauge group trace. 
The axion $\phi(x)$ field does {\it not} couple to the anomalies {\it directly}, but only through its mixing with the $b(x)$ field \eqref{mixing}.  After appropriate field redefinitions of the $b(x)$ field, 
$b(x) \, \to \, b(x) + \gamma \, \phi (x)$,  in order to diagonalise the $b(x)$- and $\phi(x)$-field  kinetic terms, and a redefinition of the $\phi(x)$ field, so that it acquires a canonically-normalised kinetic term, one obtains~\cite{mp} a coupling of the axion $\phi(x)$ field to the mixed anomalies,  
\be\label{phiff} 
\frac{\gamma}{f_b\, \sqrt{1-\gamma^2}}\, \phi(x)  \, \Big[ \mathcal C_R \, \mathbf R\, \wedge \widetilde{\mathbf R} \, - \mathcal C_F\, {\rm Tr}\Big({\mathbf F} \, \wedge \,\widetilde {\mathbf F}\Big)\Big]~.
\ee
The coefficient multiplying $\phi(x)$ in  \eqref{phiff}, outside the square brackets, defines the inverse of the axion coupling constant in this case, $f^{-1}_\phi = \frac{\gamma}{f_b \sqrt{1-\gamma^2}}$. The instantons that may characterise the non-Abelian stringy gauge group lead, then, to the formation of an axion-$\phi(x)$ potential of the form \eqref{transpl}. For a stringy-axion kinetic mixing coefficient $\gamma \sim \lambda \ll 1$, {\it i.e.} assuming that $\gamma$ is generated by the same stringy non-perturbative effects that generate the Yukawa interactions, one obtains the $f_\phi$ used above (\eqref{trans}).}
}
with $c_1$ a numerical coefficient, and 
\be\label{lamS}
\lambda \sim e^{-S_{\rm inst}} \ll 1,
\ee 
with $S_{\rm inst}$ some non-perturbative (e.g. string-instanton) action scale. On setting 
$\Lambda_{\rm QG} \sim M_P$, and 
expanding the 
arguments of the cosine in \eqref{transpl}, we obtain
\be\label{cosine}
U_{\rm string} \simeq \frac{1}{2} \, c_1^2 \, \lambda^2 \, {M_P^2}\, \phi^2 - \frac{c_1^4}{4!} \, \lambda^4 \, \phi^4 + {\mathcal O}\Big( \frac{\lambda^6}{M_P^2 }\Big) \simeq \frac{1}{2} \, c_1^2 \, \lambda^2 \, {M_P^2} \, \phi^2
\ee
for small non-perturbative Yukawa couplings $|\lambda| \ll 1$ we are considering here, which justifies keeping only a bare mass term in the action.\footnote{The reader should also notice that such potentials generate negligible quartic interactions among the pseudoscalar fields $\phi$, which however appear with the opposite sign as compared to the model of \cite{bashir}, and hence the physical implications would be different, even if we kept such axion self-interaction terms in our analysis.} The coefficient $c_1$, and hence the axion constant in this case is fixed by 
the SD treatment \eqref{SD1cb}. In fact, to be more precise, one may consider axion constants of the form $f_\phi = c_1\, M_p /{(\lambda + {\mathcal F}(\lambda)})$, where the (small) function ${\mathcal   F}(\lambda)$ 
includes non-perturbatively small $\lambda$-corrections and 
 is fixed by \eqref{SD1cb}.
 
\nicktext{Although, as mentioned in the introduction, the Yukawa-coupling-induced masses of our stringy axion refer to a different epoch of the Universe than the QCD axions, it might be worthy  
of comparing the axion masses in \eqref{SD1cb} with them \eqref{axionmass}. Taking the UV cutoff to be $M_P$, as in the model of \cite{mp}, we observe that for Yukawa couplings of the form \eqref{lamS}, 
the bare mass 
for the axion 
\be\label{baremassaxion}
M_0 \sim \frac{1}{2\pi} \, e^{-S_{\rm inst}} \, M_P \,,
\ee
whilst the dynamical (physical) 
mass, $m$, incorporating quantum effects, which is equal to the fermion mass as well,  is much smaller. If $S_{\rm inst}$ is of comparable magnitude to the instanton-action of QCD, then 
both mass scales are much smaller than the QCD axion mass \eqref{axionmass}. However, 
depending on the underlying microscopic  mechanism that generates the Yukawa coupling,
$|\lambda|$ could be potentially much larger than QCD effects,  and in this sense the bare axion mass scale could be much larger than the QCD mass.  Nonetheless, in view of its non-perturbative nature \eqref{SD1cb}, the dynamically generated axion (and fermion) mass could still be much smaller than \eqref{axionmass}.} 

\nicktext{We next remark that, although potentially present at an era of the Universe preceding the generation of any non-perturbative potential for the stringy axions, as already mentioned, our bare axion mass scale \eqref{baremassaxion} could be of comparable magnitude to the masses \eqref{axionmassstring} that could be generated by stringy instanton effects at a scale $\Lambda_i$, as per \eqref{liscales}. In the model of \cite{mp}, the KR axion constant \eqref{fb} is given by $M_P/\sqrt{3\pi}$. Taking the UV cutoff to be $M_P$, we obtain from 
\eqref{axionmassstring} and \eqref{liscales} that the stringy-instanton-induced axion mass 
would be of order  $m_a = \sqrt{3\pi} \, e^{-S_i/2}\,  M_P $, with $S_i$ an appropriate instanton action, parametrising the non-perturbative effects, which, depending on the relative magnitude of $S_{\rm inst}$ \eqref{lamS} and $S_i$, could be of comparable magnitude to $M_0$, but again the dynamical physical mass for the axion $m$ \eqref{SD1cb} is expected to be much smaller. Thus our mechanism \eqref{SD1cb} produces dynamically superlight stringy axions, and right-handed singlet fermions, at early stages of the Universe. The cosmology of such fields remains to be studied in detail, and falls beyond the scope of the present article and that of \cite{MS2}. 
The situation changes once attractive  four-fermion interactions are included in the model, as we next proceed to discuss.} 

\end{itemize}

\subsection{Inclusion of attractive four-fermion interactions \label{sec:4f}} 
\begin{figure}[t]
 \centering
  \includegraphics[clip,width=0.7\textwidth]{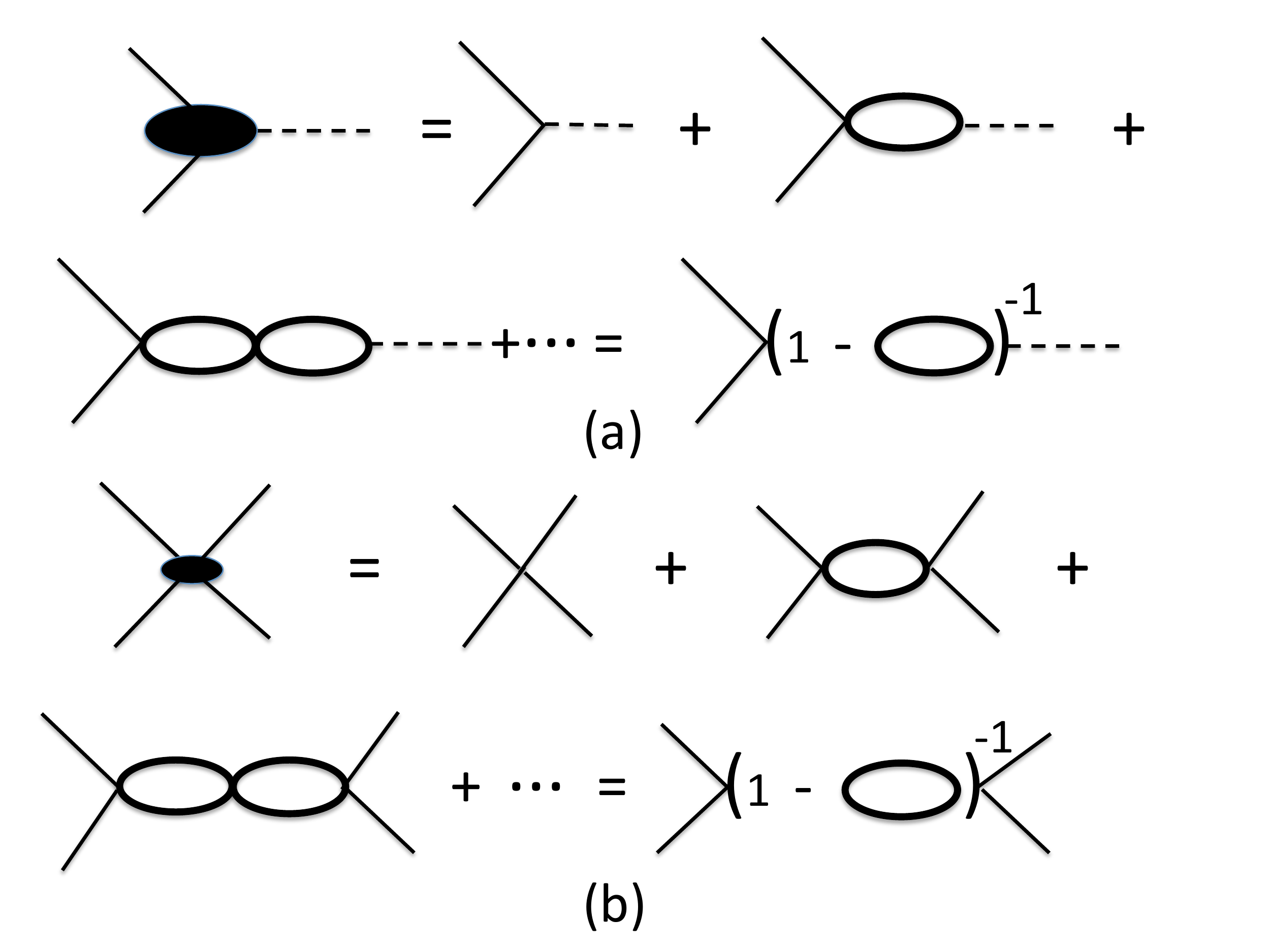} 
\caption{\it Resummation of four-fermion `bubble' diagrams (ellipse without filling in the figures) in Yukawa interaction vertex 
(upper figure (a)) and in four-fermion vertex (lower figure (b)), contributing to the leading (power law (quadratic)) UV divergences. 
Continuous lines refer to fermions, dashed lines denote (pseudo)scalar fields. The dark elliptical blob on the diagrams on the left hand 
side of the equality signs denotes schematically the result of such a resummation. }\label{fig:diagr}
\end{figure}

In this subsection, we examine the issue of dynamical mass generation in the presence of an {\it attractive} four-fermion potential term in the Lagrangian of the form
$\frac{1}{2f_4^2} \Big(\overline \psi \, \gamma^5 \, \psi \Big)^2$
so that the Lagrangian now reads:
\begin{align}
\label{4fermi}
\mathcal{L}= \frac{1}{2} \partial_{\mu} \phi \partial^{\mu} \phi +
 \overline{\psi} i \slashed{\partial} \psi + i \lambda \phi \overline{\psi}
\gamma^5 \psi - \frac{1}{2 f^2_4} (\overline{\psi} \, \gamma^5 \psi)^2,
\end{align}
where $f_4$ has mass dimension +1, and the relative negative sign between the fermion kinetic terms and the four-fermion 
interaction indicates the attractive nature of the interaction. 

To study dynamical mass generation in this case, we first linearise, as standard, the four-fermion term in (\ref{4fermi}) with the help of an auxiliary pseudoscalar field $\sigma$:
\begin{align}
\label{4fermimod}
\mathcal{L}= \frac{1}{2} \partial_{\mu} \phi \partial^{\mu} \phi + \overline{\psi}
i \slashed{\partial} \psi + i \lambda \phi \overline{\psi} \gamma^5 \psi -
\frac{f^2_4}{2} \sigma^2 -i \sigma \overline{\psi} \gamma^5 \psi.
\end{align}

The SD equations in this case are:
\begin{align}
\label{ffsdeqfer}
  G^{- 1}_f (k) - S^{- 1}_f (k) = \lambda \gamma^5 \int \frac{d^4 p}{(2
  \pi)^4} G_f (p) \Gamma^{(3)} (p,k) G_s (p - k) - \gamma^5 \int \frac{d^4 p}{(2
  \pi)^4} G_f (p) \Gamma^{(3)}_2 (p,k) G_{\sigma} (p - k),
\end{align}

\begin{align}
\label{ffsdeqsca}
  G^{- 1}_s (k) - S^{- 1}_s (k) = - \lambda \tmop{tr} \left[ \gamma^5 \int
  \frac{d^4 p}{(2 \pi)^4} G_f (p) \Gamma^{(3)} (p,k) G_f (p - k) \right],
\end{align}
where $ \Gamma^{(3)}_2 (p,k)$ denotes the vertex involving the $\sigma$ field and $G_\sigma$ stands for the $\sigma$ propagator, 
which we \nt{write as
\be\label{sigmaprop}
G_\sigma(k)=- \frac{i}{f_4^2}~. 
\ee}
We use the rainbow approximation for the vertices
\begin{align}\label{rvertices}
\Gamma^{(3)} (p,k)\simeq -\lambda\gamma^5,  \quad \Gamma^{(3)}_2 (p,k)\simeq \gamma^5~, 
\end{align}
\nt{As we explain in Appendix \ref{sec:resum}, the approximations \eqref{sigmaprop} and \eqref{rvertices}  are qualitatively (and quantitatively) sufficient,
despite the strongly-coupled nature of the four-fermion coupling, required for dynamical mass generation in this case, as we shall demonstrate below. 
In fact, \gnote{as originally described in \cite{BHL}}, resummation of one-loop fermion (``bubble'') diagram corrections (see Fig.~\ref{fig:diagr}) to the four-fermion interactions, 
associated with leading order (power law (quadratic)) UV divergences, leads simply to the use of `renormalised'  four-fermion ($f_{4R}$) and 
Yukawa ($\lambda_R)$ couplings, instead of the bare ones ($\lambda$ and $f_4$, respectively),  
which differ from the initial ones by factors of order one ({\it cf.} Eqs.~\eqref{fR}, \eqref{lR} and \eqref{Bval}): 
\gnote{
\be\label{renormcoupl}
\quad f_{4R} \sim \frac{f_4}{\sqrt2}~~, \quad \lambda_R \sim~2\lambda~. 
\ee
}
This substitution will be understood in all formulae appearing in the rest of the article, and also its sequel \cite{MS2}. 
For notational convenience, though, in what follows, we shall still use the notation
$\lambda$, $f_4$ to denote such renormalised couplings.}

With the above comments in mind, we next notice that, 
for vanishing external momenta, the SD equations read 
\begin{align}
  G^{- 1}_f (0) - S^{- 1}_f (0) = - \lambda^2 \gamma^5 \int \frac{d^4 p}{(2
  \pi)^4} G_f (p) \gamma^5 G_s (p) - \gamma^5 \int \frac{d^4 p}{(2 \pi)^4} G_f
  (p) \gamma^5 G_{\sigma} (p),
\end{align}
\begin{align}
  G^{- 1}_s (0) - S^{- 1}_s (0) =  \lambda^2 \tmop{tr} \left[ \gamma^5 \int
  \frac{d^4 p}{(2 \pi)^4} G_f (p) \gamma^5 G_f (p) \right].
\end{align}
As before, we compute the integrals using an UV cutoff $\Lambda$, and one arrives at the following system of equations 
\begin{eqnarray}\label{sd4f1}
  m + i \mu \gamma^5 = \frac{\lambda^2}{16 \pi^2} \frac{(m - i \mu \gamma^5)}{M^2 -
  m^2 - \mu^2} \left[ M^2 \ln \left( 1 + \frac{\Lambda^2}{M^2} \right) - (m^2
  + \mu^2) \ln \left( 1 + \frac{\Lambda^2}{m^2 + \mu^2} \right) \right] \nonumber\\
  +\frac{1}{16 \pi^2 f^2_4} (m - i \mu \gamma^5) \left( \Lambda^2 - (m^2 + \mu^2)
  \ln \left( 1 + \frac{\Lambda^2}{m^2 + \mu^2} \right) \right),
\end{eqnarray}

\begin{align}\label{sd4f2}
  M^2 =- \frac{\lambda^2}{4 \pi^2} \left[ \frac{(\Lambda^2 + m^2 + 3 \mu^2)
  \Lambda^2}{\Lambda^2 + m^2 + \mu^2} - (m^2 + 3 \mu^2) \ln \left( 1 +
  \frac{\Lambda^2}{m^2 + \mu^2} \right) \right].
\end{align}

For $m=\mu=0$ the only consistent solution is $M=0$, as before, given that the scalar SD equation 
is not affected by the inclusion of the four-fermion interactions.

On setting $\mu=0$, and including a bare (pseudo)scalar mass $M_0 \ne 0$, 
we seek for solutions $m \simeq M$ as in the previous section. In this case 
\eqref{sd4f1}, \eqref{sd4f2} become: 
 \begin{align}
 \label{4fmodm1}
  1 = \frac{\lambda^2}{16 \pi^2} \ln \left( 1 +
  \frac{\Lambda^2}{M^2} \right) + \frac{1}{16 \pi^2 f^2_4} \left( \Lambda^2 - m^2 \ln \left( 1
  + \frac{\Lambda^2}{m^2} \right) \right)
\end{align}

\begin{align}
 \label{4fmodm2}
  M^2 = M_0^2 - \frac{\lambda^2}{4 \pi^2} \left[ \Lambda^2 - m^2 \ln \left( 1 +
  \frac{\Lambda^2}{m^2} \right) \right].
\end{align}
On substituting \eqref{4fmodm2} into \eqref{4fmodm1}, we rewrite the above system of equations as
\begin{align}
 \label{4fmodm1b}
 m^2 \simeq M^2 &\simeq M_0^2 - 4 \, \lambda^2\, f^2_4  + \Big| \mathcal O (\lambda^4\, \ln \lambda^2) \Big|,  \nonumber \\
 M^2  & \simeq  M_0^2  -  \frac{\lambda^2\, \Lambda^2}{4\pi^2} \, > \, 0,  
 \end{align}
from which we determine the four-fermion dimensionful coupling in terms of the cut-off $\Lambda$:
\begin{align}\label{4fDir}
f_4 \simeq \frac{\Lambda}{4\pi} + \Big| \mathcal O (\lambda^2\,\ln \lambda^2) \Big|, \quad \lambda^2 \ll 1.
\end{align}
From \eqref{4fmodm1}, then we obtain, using \eqref{4fDir}:
\begin{align}\label{4fmodm1bc}
M^2 &\simeq m^2 \simeq  \lambda^2 \, f_4^2 + \mathcal O(\lambda^4 \ln \lambda^2)
= \lambda^2 \, \frac{\Lambda^2}{16\pi^2} + \mathcal O(\lambda^4\,\ln \lambda^2) \ll \Lambda^2, \nonumber \\
M_0^2 &\simeq   \lambda^2 \, \frac{5\,\Lambda^2}{16\pi^2} + \mathcal O(\lambda^4 \,\ln \lambda^2)\,.
\end{align}
It is interesting to note that the presence of four-fermion interaction implies  a {\it much larger} dynamical 
fermion mass as compared to the pure Yukawa case where 
$f_4 \to \infty$, \eqref{SD1cb}. \nicktext{As in the case examined in the previous 
subsection, the bare axion mass term $M_0$ could be thought of as the result of non-perturbative string physics deep in the transplanckian regime, generating a potential of the form \eqref{cosine}.}
This is not a surprise, due to the fact that the dimensionless four-fermion coupling $\Lambda^2/(2f_4^2) >> 1$ is very strong, as follows from \eqref{4fDir}. 

We also note that the value of the attractive four-fermion coupling $f_4$ in \eqref{4fDir} is consistent with the (strong) range of the four-fermion interactions required for dynamical mass generation in particle physics models, 
such as the Nambu-Jona-Lasinio (NJL) model~\cite{NJL} for chiral symmetry breaking.  \nt{In this respect, we should stress once more, that in our models we seek solutions to the SD equations for approximately-equal-in-magnitude dynamical masses for the fermion and pseudoscalar fields, generated primarily by the Yukawa coupling, and assisted by the four-fermion interactions. However, 
there are other solutions $m \ne M$, in cases where the fermion masses are generated by the traditional four-fermion interactions, as in the NJL model. We do not discuss this case here, as our objective is different.}

\nicktext{In cases of physical interest, like, for instance, the one studied in \cite{mp}, it is natural to assume $\Lambda \sim M_P$, and small Yukawa couplings generated by non-perturbative (instanton) effects, in which case the dynamically generated axion and right-handed fermion masses \eqref{4fmodm1bc} in that model 
are of order 
\be\label{modelmasses} 
m\simeq M \simeq \lambda\, \frac{M_P}{4\pi}\,.
\ee
Depending on the magnitude of $\lambda$, the axion mass \eqref{modelmasses} could be of comparable magnitude to axion masses generated by stringy instantons at later epochs of the Universe, but it could also be much larger. For instance, for the CPT Violating leptogenesis scenario of \cite{decesare}, in which KR gravitational axion backgrounds, undiluted from the inflationary era, play an important role~\cite{bams} in inducing lepton-asymmetric decays of the right handed fermions into (massless) standard model leptons and Higgs particles, during the radiation era, one needs typically right-handed neutrino masses of order $10^5$~GeV. In our scenario above, if such masses are generated dynamically through our Yukawa coupling mechanism, one needs $\lambda \sim 10^{-14}$, which is sufficiently small for our SD analysis to be consistent.  The magnitude of $\lambda$ is a property of a microscopic string theory and we cannot make further comments here on this important issue.} 
The example discussed in \cite{mp}  and reviewed in the introductory section \ref{sec:intro}, contains Majorana fermions, for which the above results hold qualitatively intact, as we demonstrate in section \ref{sec:maj}.

\subsection{Non-Hermitian Yukawa Interactions: dynamical (pseudo) scalar mass only}

In this subsection, we would like to discuss dynamical mass generation for anti-Hermitian Yukawa interactions, whose use has been motivated in the introduction. As we demonstrate in detail 
in Appendix \ref{sec:nhyi}, for non-Hermitian Yukawa interactions, there is no dynamical mass generation, not even for the case $m=\mu=0$, {\it i.e.} also the axion mass in that case {\it cannot} be generated dynamically. 

This is to be expected according to general energetics arguments, as discussed in  \cite{AM}, which we briefly review below. To this end, we consider a Dirac fermion and a real scalar field, for concreteness. The extension to the pseudoscalar case is immediate. The Lagrangian in this case reads:
\be\label{model}
L=\frac{1}{2}\partial_\mu\phi\partial^\mu\phi-\frac{M^2}{2}\phi^2+\bar\psi i\slashed\partial\psi-m\bar\psi\psi+\lambda\phi\bar\psi\gamma^5\psi~.
\ee
For real $\lambda$, the Yukawa interaction is imaginary, since $\bar\psi\gamma^5\psi$ is anti-Hermitian. 
The mathematical properties and consistency, as far as unitarity and Lorentz covariance properties (including those of improper Lorentz transformations) are concerned, are studied in some detail in \cite{AM}, and references therein, where we refer the interested reader for details. 

To discuss dynamical mass generation for the fermions, we first assume $m=0$ in \eqref{model} and a non-zero bare mass $M = M_0 \ne 0$ for the $\phi$ field. As discussed in Appendix \ref{sec:nhyi}, path integral quantisation of the theory requires a Euclidean formalism (with a metric signature convention (+,+,+,+)), in which the 
anti-Hermitian Yukawa interaction appears as a phase:
\be
Z_\lambda[j,\bar\eta,\eta]=\int{\cal D}[\phi,\psi,\bar\psi]\exp\left(-S_{Herm}-S_{antiHerm}-S_{sources}\right)~,
\ee
where
\bea\label{defS}
S_{Herm}&=&\int d^4x\left(\frac{1}{2}\partial_\mu\phi\partial^\mu\phi+\frac{M_0^2}{2}\phi^2+\bar\psi i\slashed\partial\psi \right), \quad
S_{sources} = \int d^4x (j\phi+\bar\eta\psi+\bar\psi\eta),\nn \\
S_{antiHerm}&=&-\lambda\int d^4x~\phi\bar\psi\gamma^5\psi~=~i\lambda\int d^4x ~\phi \, \Phi
~~~~\mbox{with}~~~~\Phi\equiv \mbox{sign}(i\bar\psi\gamma^5\psi)|\bar\psi\gamma^5\psi|~.\nonumber
\eea
From a basic property of complex calculus then we obtain the inequality
\bea\label{vwar}
\exp(-S^{ferm}_{eff})&\le&\int{\cal D}[\phi]~
\Big|\exp\left(-\int_x~\bar\psi i\slashed\partial\psi +\bar\eta\psi+\bar\psi\eta+\frac{1}{2}\phi G^{-1}\phi
+i\lambda\phi\Phi\right)\Big|\\
&=&\int{\cal D}[\phi]~\exp\left(-\int_x~\bar\psi i\slashed\partial\psi +\bar\eta\psi+\bar\psi\eta
+\frac{1}{2}\phi G^{-1}\phi\right)~,\nonumber
\eea
such that the Euclidean $S^{ferm}_{eff}$, which plays the role of {\it vacuum energy functional}, 
is {\it larger} than that for the free theory, and one cannot expect fermion dynamical mass generation~\cite{VW},  
in contrast to the usual Hermitian case, where such a dynamical mass lowers the energy of the system.
In our case this can be confirmed explicitly, by integrating out the massive scalar field to obtain
the fermionic effective action $S^{ferm}_{eff}$:
\bea
\exp(-S^{ferm}_{eff})&\equiv&\exp\left(-\int_x~\bar\psi i\slashed\partial\psi+m\bar\psi\psi+\bar\eta\psi+\bar\psi\eta\right)
\int{\cal D}[\phi]\exp\left(-\int_x\frac{1}{2}\phi G^{-1}\phi+i\lambda\phi~ \Phi\right)\\
&=&\exp\left(-\int_x~\bar\psi i\slashed\partial\psi+m\bar\psi\psi+\bar\eta\psi+\bar\psi\eta
+\frac{\lambda^2}{2}\Phi G\Phi\right)~,
\eea
where $G^{-1}=-\Box+M_0^2$ and $\Phi$ is defined in eqs.(\ref{defS}). Ignoring higher-order derivatives, 
which are not relevant for our basic arguments here, we obtain
\be\label{Sefffermion}
S^{ferm}_{eff}\simeq\int_x~\bar\psi i\slashed\partial\psi+m\bar\psi\psi+\frac{\lambda^2}{2M_0^2}|\bar\psi\gamma^5\psi|^2~,
\ee
which includes a {\it repulsive} 4-fermion interaction, and thus increases the energy of the system, as per the 
generic argument \eqref{vwar}. Hence, 
dynamical mass generation for the fermions is not possible, 
since such a process is related to the formation of an appropriate condensate that lowers the energy of the system, compared with the massless case.

We also remark that the scalar effective action $S^{scal}_{eff}$ is obtained after integrating out massive fermions 
\be
\exp(-S^{scal}_{eff})\equiv\exp\left(-\int_x\frac{1}{2}\partial_\mu\phi\partial^\mu\phi+\frac{M_0^2}{2}\phi^2+j\phi\right)
\int{\cal D}[\psi,\bar\psi]\exp\left(-\int_x\bar\psi(i\slashed\partial+m-\lambda\phi\gamma^5)\psi\right)~.\nonumber
\ee
For a constant scalar field configuration $\phi_0$, the effective potential is then~\cite{AM}
\be
U_{eff}(\phi_0)=\frac{M_0^2}{2}\phi_0^2- \mbox{Tr}\left\{\ln(\slashed p+m-\lambda\phi_0\gamma^5)\right\}~,
\ee
such that 
\be\label{dUeff}
\frac{dU_{eff}}{d\phi_0}
=M_0^2\, \phi_0 + \frac{\lambda^2\phi_0}{4\pi^2}
\left(\Lambda^2-(m^2-\lambda^2\phi_0^2)\ln\left(\frac{\Lambda^2+m^2-\lambda^2\phi_0^2}{m^2-\lambda^2\phi_0^2}\right)\right)~,
\ee
where $\Lambda$ is the UV cut off. 
The energies are therefore real, and the non-Hermitian theory is self consistent, for $m^2\ge\lambda^2\phi_0^2$, as expected from the study in \cite{BJR}. 
In the limit $\lambda^2\phi_0^2\to m^2$, the effective potential 
becomes a mass term 
\be\label{Ueff}
U_{eff}\to \frac{1}{2}(M^{(1)})^2\phi_0^2~~~~\mbox{with}~~~~(M^{(1)})^2=M_0^2+\frac{\lambda^2}{4\pi^2}\Lambda^2~.
\ee
From eq.(\ref{Ueff}) it is also clear that dynamical generation of a scalar (or axion in our case of interest) mass is
possible in the anti-Hermitian Yukawa interaction model \eqref{model}, since on setting the bare mass to zero, $M_0 =0$, one obtains from \eqref{Ueff} 
\begin{align}\label{dynam}
(M^{(1)})^2_0 = \frac{\lambda^2}{4\pi^2}\Lambda^2 > 0. 
\end{align}
The reader should notice that this is actually the result one would obtain from a one-loop calculation. 
So, in the absence of a bare scalar mass, the anti-Hermitian Yukawa interaction actually generates dynamically 
a scalar mass.
In Appendix \ref{sec:nhyi}, this conclusion is reached by a detailed analysis of the pertinent SD equations for dynamical mass generation of this non-Hermitian model.

At this point we would like to make an important remark concerning the mathematical self consistency of the non-Hermitian Yukawa model~\cite{AM}. The classical equations of motion for the (pseudo) scalar field, with a generic mass $M$,
\begin{align}\label{axioneom}
\Box \phi +M^2\phi= \lambda \, \overline \psi \gamma^5 \psi, 
\end{align}
imply, in view of the non-hermiticity of the right-hand-side, that classically, the allowed solutions are the ones
allowing a free (pseudo)scalar field and a vanishing chiral condensate for the fermions, for small in magnitude, but non-vanishing, Yukawa coupling $\lambda$:
\begin{align}\label{cond}
\Box \phi+M^2\phi=0, \qquad \langle 0 | \overline \psi \gamma^5 \psi|  0\rangle =0,
\end{align}
where we have identified classical solutions with appropriate vacuum expectation values, as standard in field theory. 

As we have discussed in Appendix \ref{sec:appA}, in this work we seek for solutions in which the chiral fermionic mass $\mu=0$, 
which is consistent with eqs.\eqref{cond}.  
 
Before closing this subsection, we also mention that, in the case of non-Hermitian Yukawa interactions, the presence of sufficiently strong four-fermion interactions, allows for dynamical mass generation, for both fermion and (pseudo)scalar fields. For details we refer the reader 
to Appendix \ref{sec:nh4f}.

\section{Schwinger-Dyson Dynamical Mass generation with Majorana fermion \label{sec:maj}}

In this section, motivated by the model of \cite{mp}, we will repeat our analysis above but for the case of Majorana fermions. Using $\psi^M = \psi_R + \psi^{C}_R$ and the chiral basis is easy to see that
\begin{align}
\overline{\psi}^{C}_R \, \psi_R - \overline{\psi}_R \, \psi^{C}_R = \overline{\psi^M} \gamma^5 \psi^M.
\end{align}
If we work in the flat space-time background, the model is of the type
\begin{align}
\label{typemodel}
\mathcal{L}= \frac{1}{2} \partial_{\mu} \phi \partial^{\mu} \phi +
\frac{1}{2} \overline{\psi^M} \, i \, \slashed{\partial} \psi^M + i \lambda \phi \overline{\psi^M}
\gamma^5 \psi^M - \frac{1}{2 f^2_4} (\overline{\psi^M} \, \gamma^5 \psi^M)^2,
\end{align}
where the reader should note the factor $1/2$ in the fermion kinetic part to avoid double counting. Moreover, from our discussion in the previous section, it becomes clear that there will be no dynamical mass generation if we only have the Yukawa interaction; hence in \eqref{typemodel} we added the appropriate chiral four-fermion {\it attractive} interaction with (dimensionful) coupling $1/(2f_4^2)$.

As in the previous section, upon using an auxiliary pseudoscalar field $\sigma$ to linearise the fermion self interaction, we arrive at the Lagrangian
\begin{align}
\label{typemodel2}
\mathcal{L}= \frac{1}{2} \partial_{\mu} \phi \partial^{\mu} \phi +
\frac{1}{2} \overline{\psi^M}\, i \, \slashed{\partial} \psi^M + i \lambda \,\phi \, \overline{\psi^M}
\gamma^5 \psi^M - \frac{f^2_4}{2} \sigma^2 - i \,\sigma \, \overline{\psi^M} \, \gamma^5 \psi^M.
\end{align}
The Schwinger-Dyson equations for the Lagrangian (\ref{typemodel2}) take the form
\begin{align}
  G^{- 1}_f (k) - S^{- 1}_f (k) = 2\lambda \gamma^5 \int \frac{d^4 p}{(2
  \pi)^4} G_f (p) \Gamma^{(3)} (p,k) G_s (p - k) - 2 \gamma^5 \int
  \frac{d^4 p}{(2 \pi)^4} G_f (p) \Gamma^{(3)}_2 (p,k) G_{\sigma} (p - k),
\end{align}

\begin{align}
  G^{- 1}_s (k) - S^{- 1}_s (k) = - \lambda \tmop{tr} \left[ \gamma^5 \int
  \frac{d^4 p}{(2 \pi)^4} G_f (p) \Gamma^{(3)} (p,k) G_f (p - k) \right],
\end{align} 
where $\Gamma^{(3)}_2 (p,k)$ denotes the vertex involving the $\sigma$ field. Using the rainbow approximation and solving the SD equations we obtain

\begin{eqnarray}
  m + i \mu \gamma^5 = \frac{\lambda^2}{8 \pi^2} \frac{(m - i \mu \gamma^5)}{M^2 -
  m^2 - \mu^2} \left[ M^2 \ln \left( 1 + \frac{\Lambda^2}{M^2} \right) - (m^2
  + \mu^2) \ln \left( 1 + \frac{\Lambda^2}{m^2 + \mu^2} \right) \right]\nonumber \\
  +
  \frac{(m - i \mu \gamma^5)}{8 \pi^2 f^2_4} \left( \Lambda^2 - (m^2 + \mu^2)
  \ln \left( 1 + \frac{\Lambda^2}{m^2 + \mu^2} \right) \right),
\end{eqnarray}

\begin{align}
  M^2 = - \frac{\lambda^2}{4 \pi^2} \left[ \frac{(\Lambda^2 + m^2 + 3 \mu^2)
  \Lambda^2}{\Lambda^2 + m^2 + \mu^2} - (m^2 + 3 \mu^2) \ln \left( 1 +
  \frac{\Lambda^2}{m^2 + \mu^2} \right) \right].
\end{align}

The structure of the equations is qualitatively similar to the Dirac fermion, and thus the solutions are similar as in that case, studied previously.

 For $m, M \ne 0$, and $\mu=0$ we have (including a bare mass $M_0$ for the scalars) 
\begin{align}\label{SDM1}
  1 = \frac{\lambda^2}{8 \pi^2} \frac{1}{M^2 - m^2} \left[ M^2 \ln \left( 1 +
  \frac{\Lambda^2}{M^2} \right) - m^2 \ln \left( 1 + \frac{\Lambda^2}{m^2}
  \right) \right] + \frac{1}{8 \pi^2 f^2_4} \left( \Lambda^2 - m^2 \ln \left(
  1 + \frac{\Lambda^2}{m^2} \right) \right),
\end{align}

\begin{align}\label{SDM2}
  M^2 =  M_0^2 -\frac{\lambda^2}{4 \pi^2} \left[ \Lambda^2 - m^2 \ln \left( 1 +
  \frac{\Lambda^2}{m^2} \right) \right].
\end{align}
We proceed as before, and we get the solutions $m \simeq M$ = finite. 
For $f_4 \to \infty$, we obtain the analogue of \eqref{SD1cb}
\bea\label{SD1cbe}
m^2 &\simeq&  \exp \Big(-\frac{8\, \pi^2}{\lambda^2}\Big) \, \Lambda^2, \quad |\lambda| \ll 1, \nonumber \\
M^2 &\simeq &  m^2 = M_0^2 - \frac{\lambda^2}{4\pi^2}\, \Lambda^2, \quad M_0^2 
= \frac{\lambda^2}{4\pi^2}\, \Lambda^2 + \exp \Big(-\frac{8\, \pi^2}{\lambda^2}\Big) \, \Lambda^2.
\eea
which is similar to the Dirac fermion case. \nt{ We remind the reader that the couplings $\lambda, f_4$ pertain to the renormalised ones \eqref{renormcoupl}, due to the strong nature of the four-fermion contact interactions.}

For $f_4$ finite, we encounter a qualitatively similar solution for fermion and scalar masses $m \simeq M$ as in 
\eqref{4fmodm1bc}, but the value of $f_4$ is now: 
\begin{align}\label{4fDirMaj}
f_4 \simeq \frac{\Lambda}{2\sqrt{2}\,\pi} + \Big| \mathcal O (\lambda^2\, \ln \lambda^2) \Big|, \quad \lambda^2 \ll 1.
\end{align}
and $M_0^2=\frac{3\, \lambda^2}{8\pi^2}\, \Lambda^2$. 

Again, as with the Dirac case, the value of the four-fermion coupling \eqref{4fDirMaj} is consistent with the (strong) coupling regime required for dynamical 
mass generation in the NJL model~\cite{NJL}. 
We also remind the reader that, in the context of the model considered in \cite{mp}, we can naturally take the UV cutoff $\Lambda \sim M_P$.

\section{Conclusions and Outlook \label{sec:concl}}

In this work we considered dynamical mass generation, \`a la Schwinger-Dyson (SD), for field theory models involving 
(pseudo)scalar fields interacting with (Dirac or Majorana) fermions via chiral Yukawa Interactions. We considered both Hermitian and non-Hermitian Yukawa couplings, in the presence, in general, of  attractive (real) four-fermion interaction terms in the Lagrangian. The presence of additional attractive four-fermion interactions might be motivated by microscopic considerations, {\it e.g.} ultra-heavy axion exchanges within an underlying string theory model, such as the one discussed in \cite{mp}, which the effective field theories we consider can be embedded into. 
In the absence of four-fermion interactions, we find that only (pseudo)scalar field mass generation was possible in the case of the non-Hermitian Yukawa interaction, but no fermion mass generation. For the Hermitian model, the situation was opposite, that is, no dynamical (pseudo)scalar mass, but dynamical fermion mass is possible, provided one adds a non-trivial bare mass for the (pseudo)scalar fields.  
On the other hand, upon the inclusion of sufficiently strong four-fermion interactions, dynamical mass generation for fermion fields (Dirac or Majorana)  is possible in both the Hermitian and non-Hermitian cases, under appropriate conditions. 

We note that the non-perturbative feature of SD equations becomes clear when one obtains a fermion dynamical mass similar to the one in eq. \eqref{SD1cb}. This solution appears naturally in the fermionic case, after dividing both sides of the SD equation by $m$. This simplification does not happen for the scalar SD equation though, where the result is similar to the one provided by a one-loop treatment. It should be stressed that the generation of a (pseudo)scalar mass in the non-Hermitian Yukawa model studied here is still dynamical, even if it is perturbative in nature.

One of the main future directions of research we plan to pursue is to consider the extension of the models presented here, in particular the non-Hermitian ones, to incorporate gravity and gravitational anomalies, as  in \cite{mp}. In the non-Hermitian case, the latter would also lead to additional non-Hermitian interactions, which might affect dynamical mass generation. \nicktext{This is considered in the sequel paper \cite{MS2}.}

In addition, as mentioned in the introduction, such (self) interactions among sterile neutrinos and axions, that could provide candidates for dark matter components in the Universe, might be useful in discussing the growth of cosmic structures and in particular the core-halo structure of galaxies, thereby offering ways to alleviate current tensions between the conventional $\Lambda$CDM-based simulations of galactic structure and observations.  Such studies can in principle provide a range of phenomenologically relevant values for the self interaction couplings and masses, and, in view of the results our current work, this can in principle constrain the parameters $\lambda, \Lambda$ used in our analysis.

\section*{Note Added in Proof}

While this article was being submitted for publication, a paper appeared~\cite{felski2}, in which dynamical mass generation for fermions is discussed in the context of a Nambu-Jona-Lasinio model with a non-Hermitian interaction with a background field $i B_\mu \bar \psi \gamma^\mu \gamma^5 \psi$. The motivation and results of our work are different from theirs, as we consider a 
non-Hermitian Yukawa interaction, leaving the four-fermion interactions Hermitian. 

\section*{Acknowledgements}
AS wishes to thank the Department of Physics of King's College London for a visiting doctoral student appointment, during which the current work was completed. 
The work  of NEM and JA is supported in part by the UK Science and Technology Facilities  research Council (STFC) under the research grants ST/P000258/1 and 
ST/T000759/1. NEM also acknowledges a scientific associateship (``\emph{Doctor Vinculado}'') at IFIC-CSIC-Valencia University, Valencia, Spain. The work of AS is supported by the CONICYT-PFCHA/Doctorado Nacional/2017-21171194.
\appendix

\section{Schwinger Dyson Equations \label{sec:appA}}

\numberwithin{equation}{section}

\setcounter{equation}{0}

In this Appendix we provide details of the construction of the SD equations pertinent to dynamical mass generation, 
in both the Hermitian  and non-Hermitian Yukawa interaction cases. 

\subsection{Resummation of fermion-loop (``Bubble diagram'') corrections to the four-fermion interactions  \label{sec:resum}}

\nt{The strong coupling nature of the four-fermion interactions encountered in our approach, requires
to take serious account of such quantum effects when formulating our SD equations for the 
four-fermion and Yukawa-coupling vertices. As we shall argue below, when concentrating on the 
power-law (quadratic) UV divergences, which suffice for our leading-order analysis of the dynamical masses, 
such quantum corrections (see fig.~\ref{fig:diagr}) can be resummed, with the result that the SD analysis based 
on the rainbow approximations for the Yukawa coupling and four-fermion vertices \eqref{sigmaprop} and \eqref{rvertices} 
(and the corresponding ones in other parts of the article) becomes qualitatively and quantitatively valid, 
if one views the couplings $\lambda$ and $f_4$, as the renormalised ones, after the resummation, which we now proceed to define.} 

\nt{To this end, we first remark that the one-loop correction to the four-fermion interaction is the ``bubble'' digaram $B$ 
for vanishing external momentum, defined as:
\be\label{B=}
\frac{i}{2f_4^2}B=-\left(\frac{i}{2f_4^2}\right)^2\mbox{Tr}\int\frac{d^4p}{(2\pi)^4}\frac{i^2}{(\slashed p-m+i\epsilon)^2}
\simeq\frac{i\Lambda^2}{16\pi^2f_4^4}~,
\ee
where we took into account the Fermi statistics of the fermion loop, and in the last equality we have reverted to Euclidean (E) momentum space, 
via $d^4 p\to i d^4 p_E$, and $p_\mu p^\mu \to p_E^2$, which properly implements the ultraviolet cutoff $\Lambda$. 
We thus obtain 
\be\label{Bexpr}
B=\frac{\Lambda^2}{8\pi^2f_4^2} > 0~.
\ee
This leads (see fig.~\ref{fig:diagr}(b)) to the following dressed (``renormalised'') four-fermion coupling, $f_{4 {\rm R}}$, 
where only the dominant contributions are kept (power-law divergences), as is sufficient for our purposes in this work: 
\be\label{fR}
\frac{i}{2f^2_{4 {\rm R}}}= \frac{i}{2f_4^2}(1+B+B^2+\cdots)=\frac{i}{2f_4^2}\frac{1}{1-B}~ \quad \Rightarrow \quad f_{4R}= \sqrt{(1-B)}\, f_4~,
\ee
provided $0 < B <1$, which has to be checked {\it a posteriori} for self consistency.}

\gnote{Loop corrections to the axion-Yukawa coupling arising from $\lambda$ are negligible, since they are logarithmic and also $\lambda<<1$. 
But corrections arising from the four-fermion interaction are not negligible (see fig.~\ref{fig:diagr}(a)), 
and they also involve the resummation of bubbles $B$. This leads to the dressed (`renormalised') Yukawa coupling, $\lambda_R$:
\be\label{lR}
i\lambda_R \gamma^5 = i\lambda\gamma^5(1+B+B^2+\cdots)=\frac{i\lambda\gamma^5}{1-B} \quad \Rightarrow \quad \lambda_R 
= \frac{1}{1-B}\, \lambda~, \quad 0 < B < 1~.
\ee
Taking into account these corrections, and noting that the Schwinger-Dyson equations (\ref{ffsdeqfer}) and (\ref{ffsdeqsca}) 
involve one bare coupling $\lambda$ and one dressed coupling $\lambda_R$, we observe that Eqs.~\eqref{4fmodm1b} become
\bea
m^2&\simeq& M_0^2-4\lambda^2f_4^2~(1-B)\\
M^2&\simeq& M_0^2-\lambda^2\frac{\Lambda^2}{4\pi^2}~\frac{1}{1-B}~\nonumber
\eea
and the identification $m^2\simeq M^2$ leads to
\be\label{fL}
\frac{\Lambda^2}{f_4^2}=16\pi^2(1-B)^2~, \quad 0 < B < 1~.
\ee
From Eq.~\eqref{Bexpr} we therefore obtain
\be
B=2(1-B)^2~,
\ee
which has a consistent solution ($0 < B < 1$): 
\be\label{Bval}
B= \frac{1}{2}~.
\ee
On account of \eqref{fL}, this leads to a bare four-fermion coupling
\be
f_4=\frac{\Lambda}{2\pi}~,
\ee
and from \eqref{fR} and \eqref{lR}, we obtain for the renormalised four-fermion and Yukawa couplings
\be\label{renormc}
f_{4R} = \frac{f_4}{\sqrt2}~~, \quad \lambda_R = 2 \lambda~, 
\ee
respectively. Note that another interpretation is possible, where one considers the original Lagrangian as effective, with the coupling
$\lambda$ replaced by the dressed coupling $\lambda_R$, as a consequence of four-fermion dressing. 
One can then use the rainbow approximation, where every factor $\lambda$ is replaced by $\lambda_R$. This procedure  
corresponds to a different resummation in the Schwinger-Dyson context, but  
leads to dressed couplings $f_{4R}$ and $\lambda_R$ which are of the same order of magnitude as the results (\ref{renormc}).
}

\nt{Thus we see that the resummation of the higher-order four-fermion interactions, yield 
the same results in order of magnitude as  in the naive approach
which such interactions are ignored. Nonetheless, as we have discussed in the text, the 
result of the resummation is accounted for by considering the $\lambda$, $f_4$ couplings appearing in the 
Schwinger-Dyson analysis as the `renormalised'  ones \eqref{renormc}. 
This will be understood throughout this work, and its sequel \cite{MS2}, and also in the subsequent subsections of the Appendix.}

\subsection{Hermitian Yukawa Interactions \label{sec:hyi}}

In this case, the generating functional  is given by
\begin{align}
\label{hermitgenfun}
  Z [J, \eta, \overline{\eta}] = \int \mathcal{D} [\phi \psi \overline{\psi}] \exp
  \left\{ i \int d^4 x \left[ \frac{1}{2} \partial_{\mu} \phi \partial^{\mu}
  \phi + \overline{\psi} i \slashed{\partial} \psi + i \lambda \phi \overline{\psi} \gamma^5
  \psi \right] + i \int d^4 x [J \phi + \overline{\psi} \eta + \overline{\eta} \psi]
  \right\}.
\end{align}

First, we start with the scalar field. The equation is given by

\begin{align}
  \int \mathcal{D} [\phi \psi \overline{\psi}] \left( J (x) -
  \partial^2 \phi (x) + i \lambda \overline{\psi} (x) \gamma^5 \psi (x) \right)
  e^{i \tilde \sigma} = 0,
\end{align}
where
\begin{align}\label{equs}
\tilde \sigma \equiv \int d^4 x \left[ - \frac{1}{2} \phi \partial^2 \phi + \overline{\psi} i
\slashed{\partial} \psi + i \lambda \phi \overline{\psi} \gamma^5 \psi \right] + \int
d^4 x [J \phi + \overline{\psi} \eta + \overline{\eta} \psi]
\end{align}
Following the notation and conventions in \cite{Peskin:1995ev}, and using the generating functional for connected diagrams $W =  i \ln Z$, we easily arrive at 
\begin{align}
\label{derj}
  &\delta^{(4)} (x - y)  + \partial^2 \left( \frac{\delta^2 W}{\delta J (y)
  \delta J (x)} \right) - i \lambda \left( \frac{\delta^2 W}{\delta J (y)
  \delta \eta (x)} \right) \gamma^5 \left( \frac{\delta W}{\delta \overline{\eta}
  (x)} \right) - i \lambda \left( \frac{\delta W}{\delta \eta (x)} \right)
  \gamma^5 \left( \frac{\delta^2 W}{\delta J (y) \delta \overline{\eta} (x)}
  \right) \nonumber \\ &+ \lambda \tmop{tr} \left[ \gamma^5 \frac{\delta^3 W}{\delta J (y)
  \delta \overline{\eta} (x) \delta \eta (x)} \right] = 0
\end{align}

Defining the Legendre transform of $W$ as
\begin{align}
  W [J, \eta, \overline{\eta}] = - \Gamma [\phi \psi \overline{\psi}] - \int d^4 x [J
  \phi + \overline{\psi} \eta + \overline{\eta} \psi]
\end{align}
and following standard functional techniques~\cite{Peskin:1995ev}, 
we obtain from \eqref{derj} the following Schwinger-Dyson (SD) equation for the scalar ($s$)  propagator:
\begin{align}
  G^{- 1}_s (x - y) - S^{- 1}_s (x - y) + \lambda \tmop{tr} \left[ \gamma^5 \int d^4 v d^4 w G_f (x - v) \Gamma^{(3)} (y, w, v) G_f (w - x) \right] = 0,
\end{align}
where 
\begin{align}
\Gamma^{(3)} (z, w, v)=\frac{i \delta^3 \Gamma}{\delta \phi (z) \delta \psi (w) \delta \overline{\psi}(v)}.
\end{align}
is the vertex function of the Yukawa interaction, and we have used $S^{- 1}_s (x - y)=i\partial^2$. Passing onto Fourier space we obtain
\begin{align}
\label{sdscal}
  G^{- 1}_s (k) - S^{- 1}_s (k) = - \lambda \tmop{tr} \left[ \gamma^5 \int  \frac{d^4 p}{(2 \pi)^4} G_f (p) \Gamma^{(3)} (p, k) G_f(p - k) \right]~,
\end{align}
with the bare inverse scalar propagator being given by $S_s^{-1}(k)=-i k^2$.

To arrive at the SD equations for the fermion ($f$) propagator $G_f^{-1} (x-y)$ we proceed in a similar way. The analogue of \eqref{equs}  is now given by
\begin{align}
  \int \mathcal{D} [\phi \psi \overline{\psi}] \left( \eta (x) + i \slashed{\partial}
  \psi (x) + i \lambda \phi (x) \gamma^5 \psi (x) \right) e^{i \tilde \sigma} = 0
\end{align}
Following the same steps as for the scalar field, we obtain in Fourier space the equation
\begin{align}
\label{sdferm}
  G^{- 1}_f (k) - S^{- 1}_f (k) = \lambda \gamma^5 \left( \int \frac{d^4 p}{(2\pi)^4} G_f (p) \Gamma^{(3)} (p, k) G_s (p - k) \right)
\end{align}
with the bare inverse fermion propagator $S_f^{-1}=-i\slashed{k}$.

In our approach, we assume $|\lambda| \ll 1$, and thus we shall employ the {\it rainbow} approximation, in which the vertex does not receive
corrections: 
\begin{align}\label{rainvert}\Gamma^{(3)} \simeq - \lambda \gamma^5.~ 
\end{align}

For vanishing external
momenta the equations (\ref{sdscal}) and (\ref{sdferm}) read

\begin{align}
\label{sd2}
  G^{- 1}_s (0) - S^{- 1}_s (0) =  \lambda^2 \tmop{tr} \left[ \gamma^5 \int \frac{d^4 p}{(2 \pi)^4} G_f (p) \gamma^5 G_f (p)\right]
\end{align}
\begin{align}
\label{sd1}
  G^{- 1}_f (0) - S^{- 1}_f (0) = - \lambda^2 \gamma^5 \left( \int \frac{d^4 p}{(2 \pi)^4} G_f (p) \gamma^5 G_s (p) \right)
\end{align}

To solve (\ref{sd1}) and (\ref{sd2}) we introduce the dressed inverse propagators $G_f^{-1} (p)=-i(\slashed{p}-m-i\mu \gamma^5)$, $G_s^{-1}(p)=-i(p^2-M^2)$. Performing the momentum integration using an UV cutoff $\Lambda$, we arrive at the SD equations:

\begin{align}\label{hSD2}
  M^2 = -\frac{\lambda^2}{4 \pi^2} \left[ \frac{(\Lambda^2 + m^2 + 3 \mu^2)
  \Lambda^2}{\Lambda^2 + m^2 + \mu^2} - (m^2 + 3 \mu^2) \ln \left( 1 +
  \frac{\Lambda^2}{m^2 + \mu^2} \right) \right]
\end{align}
\begin{align}\label{hSD1}
  m + i \mu \gamma^5 = \frac{\lambda^2}{16 \pi^2} \frac{(m - i \mu \gamma^5)}{M^2 -
  m^2 - \mu^2} \left[ M^2 \ln \left( 1 + \frac{\Lambda^2}{M^2} \right) - (m^2
  + \mu^2) \ln \left( 1 + \frac{\Lambda^2}{m^2 + \mu^2} \right) \right]
\end{align}

We discuss solutions of these equations in the main text.

\subsection{Non-Hermitian Yukawa Interactions \label{sec:nhyi}}
 
The Euclidean (``$E$ '') generating functional for the non-Hermitian Yukawa interactions, 
to be used in our study of dynamical mass generation, is (the Euclidean metric signature convention is (+,+,+,+))
\begin{align}\label{partf}
  Z^E [J, \eta, \overline{\eta}] = \int \mathcal{D} [\phi  \psi
  \overline{\psi}] \exp \left\{ - \int d^4 x  \left[ \frac{1}{2} \, \partial_{\mu} \phi
  \partial^{\mu} \phi  + \overline{\psi} i\slashed{\partial} \psi - \lambda
  \phi \,  \overline{\psi} \gamma^5 \psi \right] - \int d^4 x [J
  \phi  + \overline{\psi} \eta + \overline{\eta} \psi] \right\}.
\end{align}

The analysis for the construction of the SD equations is similar to the Hermitian case. We start with

\begin{align}
  \int \mathcal{D} [\phi \psi \overline{\psi}] (J (x) - \partial^2 \phi (x) -
  \lambda \overline{\psi} (x) \gamma^5 \psi (x)) e^{-\tilde \sigma_{\tt NH}} = 0,
\end{align}

where

\begin{align}
\tilde \sigma_{\tt NH} =  \int d^4 x \left[ - \frac{1}{2} \phi \partial^2 \phi + \overline{\psi}
i \slashed{\partial} \psi - \lambda \phi \overline{\psi} \gamma^5 \psi \right] + \int d^4
x [J \phi + \overline{\psi} \eta + \overline{\eta} \psi]
\end{align}

Proceeding as in the Hermitian case, we arrive at the following equations for the scalar ($s$) and fermion ($f$) propagators in Fourier space:

\begin{align}
  G^{- 1}_s (k) - S^{- 1}_s (k) =  \lambda \tmop{tr} \left[ \gamma^5 \int \frac{d^4 p}{(2 \pi)^4} G_f (p) \Gamma^{(3)} (p, k) G_f (p - k) \right]
\end{align}
\begin{align}
  G^{- 1}_f (k) - S^{- 1}_f (k) = - \lambda \gamma^5 \left( \int \frac{d^4 p}{(2 \pi)^4} G_f (p) \Gamma^{(3)} (p, k) G_s (p - k) \right)
\end{align}
where the vertex function in the rainbow approximation in the non-Hermitian case reads:
\begin{align}\label{rainvertNH}
\Gamma^{(3)} (p,k) \simeq \lambda \gamma^5~.
\end{align}
On assuming vanishing external momenta, we may write the SD equations as:
\begin{align}
\label{sd2euc}
  G^{- 1}_s (0) - S^{- 1}_s (0) =  \lambda^2 \tmop{tr} \left[ \gamma^5 \int \frac{d^4 p}{(2 \pi)^4} G_f (p) \gamma^5 G_f (p) \right]
\end{align}
\begin{align}
\label{sd1euc}
 G^{- 1}_f (0) - S^{- 1}_f (0) = - \lambda^2 \gamma^5 \left( \int \frac{d^4 p}{(2 \pi)^4} G_f (p) \gamma^5 G_s (p) \right)
\end{align}

We use the expressions for the dressed inverse propagators $G_f^{-1}(p)=\slashed{p}+m+\mu \gamma^5$, $G_s^{-1}(p)=p^2+M^2$ and for the bare inverse propagators $S_f^{-1}(p)=\slashed{p}$, $S_s^{-1}(p)=p^2$. 
We restrict ourselves to the case where $\mu $ is real, $\mu \in {\tt R}$. This stems from the fact that we are interested in the (physically relevant) case where the energies of the system in the massive phase are real,  for which one must have the following condition among the (dynamically generated) mass parameters~\cite{BJR,AM}
\begin{align}\label{mumrel}
|\mu| \le |m|.
\end{align}
For completeness, we mention that the latter condition also guarantees unitarity, in the sense that the respective probability of the non-Hermitian fermionic subsystem is less than unity, and thus well defined, and conserved~\cite{AMS}.

Performing the 
Euclidean momentum integrations with an UV cutoff $\lambda$, we obtain

\begin{align}
\label{sdnhb}
  M^2 =  \frac{\lambda^2}{4 \pi^2} \left[ \frac{(\Lambda^2 + m^2 - 3 \mu^2) \Lambda^2}{\Lambda^2 + m^2 - \mu^2} - (m^2 - 3 \mu^2) \ln \left( 1 + \frac{\Lambda^2}{m^2 - \mu^2} \right) \right]
\end{align}
\begin{align}
\label{sdnha}
  m + \mu \gamma^5 = - \frac{\lambda^2}{16 \pi^2} \frac{(m - \mu \gamma^5)}{M^2 - m^2 + \mu^2} \left[ M^2 \ln \left( 1 +
  \frac{\Lambda^2}{M^2} \right) - (m^2 - \mu^2) \ln \left( 1 + \frac{\Lambda^2}{m^2 - \mu^2} \right) \right]
\end{align}

On setting $m=\mu=0$, we observe that there is now a consistent solution for the dynamically generated axion mass $M$, since \eqref{sdnhb} leads to
\begin{align}\label{scalar}
M^2 =  \frac{\lambda^2}{4 \pi^2} \Lambda^2 \, \ll \Lambda^2, \quad \lambda^2 \ll 1.
\end{align}

On account of \eqref{mumrel}, the case where $m =0$ but $\mu \ne 0$ is not allowed, as it would lead to unphysical situations. 

Considering $\mu=0$, we have the following system of SD equations

\begin{align}
\label{sd2mu0nh}
  M^2 =  \frac{\lambda^2}{4 \pi^2} \left[ \Lambda^2 - m^2 \ln \left( 1 + \frac{\Lambda^2}{m^2} \right) \right]
\end{align}
\begin{align}
\label{sd1mu0nh}
  1 = - \frac{\lambda^2}{16 \pi^2} \frac{1}{M^2 - m^2} \left[ M^2 \ln \left( 1 + \frac{\Lambda^2}{M^2} \right) - m^2 \ln \left( 1 + \frac{\Lambda^2}{m^2} \right) \right]
\end{align}

We will now consider solutions $m \simeq M \ll \Lambda$. From (\ref{sd1mu0nh}):
\begin{align}
  - \frac{16 \pi^2}{\lambda^2} = \ln \left( 1 + \frac{\Lambda^2}{M^2} \right), 
\end{align}
which is inconsistent. It is also readily seen that the case 
$\mu=M=0$ also does not lead to fermion mass generation.

Hence dynamical fermion mass is not possible for pure anti-Hermitian Yukawa interactions, only scalar mass can be generated
dynamically. This was discussed in the text, where generic energetics arguments were provided to support the above results.

\subsection{Non-Hermitian Yukawa Interactions in the presence of Attractive Four-Fermion interactions \label{sec:nh4f}}

We next proceed to discuss explicitly such extra four-fermion interactions for the non-Hermitian case. 

In this case we consider the Euclidean action
\begin{align}\label{part4f}
  Z^E [J, \eta, \overline{\eta}] = \int \mathcal{D} [\phi  \psi
  \overline{\psi}] \exp \left\{ - \int d^4 x  \left[ \frac{1}{2} \, \partial_{\mu} \phi
  \partial^{\mu} \phi  + \overline{\psi} i\slashed{\partial} \psi - \lambda
  \phi \,  \overline{\psi} \gamma^5 \psi  + \frac{1}{2 f^2_4} (\overline{\psi} \, \gamma^5 \psi)^2\right] - \int d^4 x [J
  \phi  + \overline{\psi} \eta + \overline{\eta} \psi] \right\}.
\end{align}
The reader should have noticed that the attractive four-fermion interactions come with the same sign as the 
kinetic axion term and the fermion term, given that in the Euclidean formalism the integrand of the exponent of the partition function is actually the Hamiltonian of the model.

We can rewrite the action (\ref{part4f}) using an auxiliary field $\sigma$ as {\bf } 
{\small \begin{align}
  Z^E [K, J, \eta, \overline{\eta}] = \int \mathcal{D} [\sigma \phi \psi \overline{\psi}] \exp \left\{ - \int d^4 x \left[ \frac{1}{2} \partial_{\mu} \phi \partial^{\mu} \phi + \overline{\psi} i \slashed{\partial} \psi - \lambda \phi \overline{\psi} \gamma^5 \psi + \frac{f^2_4}{2} \sigma^2  + i \, \sigma \overline{\psi} \gamma^5 \psi \right]  - \int d^4 x [J \phi  + \overline{\psi} \eta + \overline{\eta} \psi + K \sigma] \right\}
\end{align}}

The additional ingredient with  respect to the analysis in the previous section is the $\sigma$ propagator, which is given by

\begin{equation}
\label{sigmapropnh}
G_\sigma (x - y)  = \frac{\delta^2 W[J]}{\delta K (x) \delta K (y)} =  \frac{1}{f_4^2}.
\end{equation}

The computation of the SD equation for the scalar is the same as before, since there is no new term involving $\phi$. However, the fermion equation is modified by the four-fermion-interaction term. Following similar steps as above, we then easily arrive, under the rainbow approximation for the vertices, 
at the SD equations with vanishing external momenta:

\begin{equation}
  G^{- 1}_f (0) - S^{- 1}_f (0) = - \lambda^2 \gamma^5 \left( \int \frac{d^4 p}{(2 \pi)^4} G_f (p) \gamma^5 G_s (p) \right) - \gamma^5 \left( \int \frac{d^4 p}{(2 \pi)^4} G_f (p) \gamma^5 G_{\sigma} (p) \right)
\end{equation}
\begin{equation}
  G^{- 1}_s (0) - S^{- 1}_s (0) =  \lambda^2 \tmop{tr} \left[ \gamma^5 \int \frac{d^4 p}{(2 \pi)^4} G_f (p) \gamma^5 G_f (p) \right]
\end{equation}
Using the dressed inverse propagators for the fermions $G_f^{-1}(p)=\slashed{p}+m+\mu \gamma^5$, with $\mu \in {\tt R}$, under the condition \eqref{mumrel},  and the scalars $G_s^{-1}(p)=p^2+M^2$, as well as the bare inverse propagators $S_f^{-1}(p)=\slashed{k}$, $S_s^{-1}(p)=k^2$, together with $G_\sigma(p)=1/f_4^2$, 
and performing the integrals using an UV cutoff $\Lambda$, we arrive at: {\bf  }

\begin{eqnarray}
  m + \mu \gamma^5 = - \frac{\lambda^2}{16 \pi^2} \frac{(m - \mu \gamma^5)}{M^2 - m^2 + \mu^2} \left[ M^2 \ln \left( 1 +
  \frac{\Lambda^2}{M^2} \right) - (m^2 - \mu^2) \ln \left( 1 + \frac{\Lambda^2}{m^2 - \mu^2} \right) \right] \nonumber\\
  + \frac{(m - \mu \gamma^5)}{16\pi^2 f_4^2} \left( \Lambda^2 - (m^2 - \mu^2) \ln \left( 1 + \frac{\Lambda^2}{m^2 - \mu^2} \right) \right)
\end{eqnarray}

\begin{equation}\label{massMnh}
  M^2 =  \frac{\lambda^2}{4 \pi^2} \left[ \frac{(\Lambda^2 + m^2 - 3 \mu^2) \Lambda^2}{\Lambda^2 + m^2 - \mu^2} - (m^2 - 3 \mu^2) \ln \left( 1 + \frac{\Lambda^2}{m^2 - \mu^2} \right) \right]
\end{equation}

As the SD scalar mass equation is independent of $f_4$, it is straightforward to see from \eqref{massMnh} that, for $m=\mu=0$, one obtains 
the dynamically generated (pseudo)scalar mass \eqref{scalar}.

If we consider $\mu=0$ but $ m, M \ne 0$, the system of SD equations reads

\begin{equation}
\label{nh4mu0a}
  1 = - \frac{\lambda^2}{16 \pi^2} \frac{1}{M^2 - m^2} \left[ M^2 \ln \left( 1 + \frac{\Lambda^2}{M^2} \right) - m^2 \ln \left( 1 + \frac{\Lambda^2}{m^2}\right) \right] + \frac{1}{16 \pi^2 f_4^2} \left( \Lambda^2 - m^2 \ln \left(1 + \frac{\Lambda^2}{m^2} \right) \right)
\end{equation}

\begin{align}
\label{nh4mu0b1}
  M^2 =  \frac{\lambda^2}{4 \pi^2} \left[ \Lambda^2 - m^2 \ln \left( 1 + \frac{\Lambda^2}{m^2} \right) \right], 
\end{align}
which for $\Lambda \, \gg \, m \simeq M  \ne 0$, becomes
 \begin{align}
\label{nh4mu0ab}
  1 =& - \frac{\lambda^2}{16 \pi^2}  \ln \left( 1 + \frac{\Lambda^2}{M^2} \right)  + \frac{m^2}{4\, f_4^2 \, \lambda^2}  \, \Rightarrow \, m^2 \simeq M^2 \simeq 4 \, \lambda^2 \, f_4^2  +  \Big | \mathcal O (\lambda^4 \, \ln \lambda^2) \Big |~,  
 \nonumber \\
  M^2 &\simeq m^2  \simeq   \frac{\lambda^2}{4 \pi^2} \Lambda^2~,
\end{align}
which imply that 
\be
f_4 \simeq \frac{\Lambda}{4\pi} -  \Big | \mathcal O (\lambda^2\, \ln \lambda^2) \Big |, \quad \lambda^2 \ll 1.
\ee
Thus, in non-Hermitian Yukawa interactions, upon the inclusion of sufficiently strong four-fermion attractive interactions, one can obtain dynamical fermion and (pseudo) scalar mass generation.

\end{document}